\documentclass[preprint,3p]{elsarticle}

\usepackage{amssymb}

\usepackage{lineno}
\usepackage{changepage} 

\usepackage{graphicx}
\graphicspath{{figures}}
\usepackage{xcolor}
\usepackage[english]{babel}
\usepackage{multirow}

\usepackage{multicol}
\usepackage{threeparttable}
\usepackage{booktabs}
\usepackage{longtable}
\usepackage{hyperref}
\usepackage{amsthm, amsmath}
\usepackage{algorithm}
\usepackage{algpseudocode}
\usepackage{placeins}
\usepackage{soul}
\usepackage{listings}

\usepackage{bm}
\usepackage{bbm}
\usepackage{mathtools}

\definecolor{codegreen}{rgb}{0,0.6,0}
\definecolor{codegray}{rgb}{0.5,0.5,0.5}
\definecolor{codepurple}{rgb}{0.58,0,0.82}
\definecolor{backcolour}{rgb}{0.95,0.95,0.92}

\lstdefinestyle{mystyle}{
    backgroundcolor=\color{backcolour},   
    commentstyle=\color{codegreen},
    keywordstyle=\color{magenta},
    numberstyle=\tiny\color{codegray},
    stringstyle=\color{codepurple},
    basicstyle=\ttfamily\footnotesize,
    breakatwhitespace=false,         
    breaklines=true,                 
    captionpos=b,                    
    keepspaces=true,                 
    numbers=left,                    
    numbersep=5pt,                  
    showspaces=false,                
    showstringspaces=false,
    showtabs=false,                  
    tabsize=2
}

\usepackage{amsfonts}
\usepackage{subcaption}

\newcommand{\beq}{\begin{equation}}
\newcommand{\eeq}{\end{equation}}

\def\ten   #1{\boldsymbol{#1}}
\def\vec   #1{\boldsymbol{#1}}
\journal{arXiv}

\newtheorem*{remark}{Remark}
\newcommand{\alg}{Alg.\,}

\newcommand{\eq}{}
\newcommand{\eqs}{}
\newcommand{\fig}{Fig.\,}

\newcommand{\sect}{Section\,}
\newcommand{\sects}{Sections\,}
\newcommand{\tab}{Tab.\,}

\renewcommand{\vec}[1]{\bm{#1}} % Second order tensors are bold and upright
\newcommand{\tenII}[1]{\bm{#1}} % Second order tensors are bold and upright
\newcommand{\Space}[1]{\mathbbm{#1}} % Spaces are blackboard bold
\newcommand{\I}{\tenII{I}} % Identity tensor
\newcommand{\brac}[1]{\left[#1\right]}  % Brackets
\newcommand{\fOf}[1]{\left(#1\right)}  % Function of
\newcommand{\set}[1]{\left\{#1\right\}}  % Set
\newcommand{\tr}[1]{\text{tr}\fOf{#1}}  % Trace
\renewcommand{\det}[1]{\text{det}\fOf{#1}}  % Determinant
\newcommand{\pd}[2]{\frac{\partial #1}{\partial #2}}  % partial derivative
\newcommand{\pdd}[3]{\frac{\partial^2 #1}{\partial #2\partial #3}}  % second partial derivative
\DeclareMathOperator*{\argmin}{arg\,min}

\renewcommand{\exp}[1]{\text{exp}\fOf{#1}}
\newcommand{\plaind}{\text{d}}

\newcommand{\sign}[1]{\text{sgn}\fOf{#1}}

\newcommand{\sym}[1]{\text{sym}\fOf{#1}}

\newcommand{\z}{\bm{z}}

\newcommand{\stimes}{\overline{\underline{\otimes}}}

\newcommand{\OO}{\mathbbm{O}}
\newcommand{\cc}{\mathbbm{c}}
\newcommand{\CC}{\mathbbm{C}}

\newcommand{\GG}{\mathbbm{G}}

\newcommand{\RR}{\Space{R}}

\newcommand{\matVal}[1]{\uppercase{#1}} % Material values are capital
\newcommand{\spaVal}[1]{\lowercase{#1}} % Spatial values are lower case
\newcommand{\matVec}[1]{\vec{\matVal{#1}}} % Material vectors
\newcommand{\spaVec}[1]{\vec{\spaVal{#1}}} % Spatial vectors
\newcommand{\iso}[1]{\tilde{#1}}
\newcommand{\vol}[1]{\bar{#1}}
\newcommand{\F}{\tenII{F}}
\newcommand{\nsv}{n_{\text{sv}}}
\newcommand{\StrucVec}{\matVec{a}}
\newcommand{\strucVec}{\spaVec{a}}
\newcommand{\StrucVeci}{\StrucVec^{i}}
\newcommand{\StrucVecj}{\StrucVec^{j}}
\newcommand{\strucVeci}{\strucVec^{i}}
\newcommand{\strucVecj}{\strucVec^{j}}
\renewcommand{\a}{\spaVec{a}}

\renewcommand{\b}{\bm{b}}
\renewcommand{\d}{\bm{d}}
\newcommand{\e}{\bm{e}}

\renewcommand{\u}{\bm{u}}

\newcommand{\x}{\bm{x}}
\newcommand{\btau}{\bm{\tau}}
\newcommand{\ta}{\bm{\tau}}
\newcommand{\Aset}{\mathcal{A}}
\newcommand{\A}{\tenII{A}}

\newcommand{\B}{\tenII{B}}

\newcommand{\C}{\tenII{C}}

\newcommand{\G}{\tenII{G}}

\renewcommand{\P}{\tenII{P}}

\renewcommand{\S}{\tenII{S}}

\newcommand{\X}{\vec{\matVal{x}}}
\renewcommand{\r}{\bm{r}}

\newcommand{\en}{\Psi}

\newcommand{\GIso}{\iso{\G}}

\newcommand{\GGIso}{\iso{\GG}}
\newcommand{\BIso}{\iso{\B}}

\newcommand{\FIso}{\iso{\F}}
\newcommand{\FVol}{\vol{\F}}

\newcommand{\strucVecIso}{\iso{\strucVec}}

\newcommand{\strucVecIsoi}{\strucVecIso^{i}}
\newcommand{\strucVecIsoj}{\strucVecIso^{j}}

\renewcommand{\Im}{I_m}

\newcommand{\eIso}{\alpha}

\newcommand{\eIsom}{\eIso_m}

\newcommand{\invSet}{\mathcal{I}}
\newcommand{\invSetIso}{\iso{\invSet}}
\newcommand{\iI}{I_1}
\newcommand{\iII}{I_2}
\newcommand{\iIII}{I_3}

\newcommand{\iIVind}{I_{4,ij}}
\newcommand{\iVind}{I_{5,ij}}

\newcommand{\iIi}{\iso{\iI}}
\newcommand{\iIIi}{\iso{\iII}}

\newcommand{\IThree}{I_3}

\newcommand{\ImIso}{\iso{I}_m}

\newcommand{\dom}{\Omega}
\newcommand{\domO}{\dom_{0}}

\newcommand{\boundN}{\Gamma_N}
\newcommand{\intOn}[2]{\int_{#2}#1\, \plaind #2}

\newcommand{\intOnDomO}[1]{\int_{\domO}#1\, \plaind \domO}

\newcommand{\intOnBoundN}[1]{\intOn{#1}{\boundN}}

\newcommand{\grad}[1]{\nabla_{\x}#1}
\newcommand{\Grad}[1]{\nabla_{\X}#1}

\newcommand{\vecArray}[1]{\left\{#1\right\}}
\newcommand{\act}[1]{\mathcal{#1}}
\renewcommand{\z}{\vec{z}}
\newcommand{\bfz}{\vec{z}}
\newcommand{\my}{\vec{y}}

\newcommand{\bias}{\vec{c}}
\newcommand{\spli}{s}
\newcommand{\calQ}{\mathcal{Q}}
\newcommand{\params}{\calQ}
\newcommand{\actF}{\act{F}}
\newcommand{\actFof}[1]{\actF\fOf{#1}}

\newcommand{\lay}[1]{^{(#1)}}
\newcommand{\nnInput}{\actK}
\newcommand{\actK}{\act{K}}
\newcommand{\actKof}[1]{\actK\fOf{#1}}
\newcommand{\actN}{\act{N}}
\newcommand{\actNof}[1]{\actN\fOf{#1}}
\newcommand{\shapeFn}{\phi}

\newcommand{\shapeFnI}{\shapeFn^{I}}
\newcommand{\shapeFnJ}{\shapeFn^{J}}
\newcommand{\qp}{\text{q}}
\newcommand{\pts}{\text{pts}}
\newcommand{\nqp}{n_{\qp}}
\newcommand{\npts}{n_{\pts}}
\newcommand{\dofs}{\text{dofs}}
\newcommand{\nDofs}{{n_{\dofs}}}
\newcommand{\nodes}{\text{nodes}}
\newcommand{\nNodes}{{n_{\nodes}}}

\newcommand{\el}{\text{e}}
\newcommand{\nEl}{{n_{\el}}}
\newcommand{\batch}{\text{b}}
\newcommand{\nBatch}{{n_{\batch}}}
\newcommand{\vF}{\vecArray{\F}}
\newcommand{\ven}{\vecArray{\en}}
\newcommand{\vtau}{\vecArray{\btau}}
\newcommand{\vcc}{\vecArray{\cc}}
\newcommand{\K}{\ten{K}}
\newcommand{\dd}{\Delta\d}

\newcommand{\bigO}{\mathcal{O}}

\newcommand{\questionColor}[1]{\textcolor{ctpGreen}{#1}}
\newcommand{\RV}[1]{#1}

\newcounter{example}

\begin{document}

\begin{frontmatter}

	\title{COMMET: orders-of-magnitude speed-up in finite element method via batch-vectorized neural constitutive updates}

	\author[inst1,inst2]{Benjamin Alheit}
	\author[inst1]{Mathias Peirlinck\corref{cor1}}
	\ead{mplab-me@tudelft.nl}
	\author[inst2]{Siddhant Kumar\corref{cor1}}
	\ead{sid.kumar@tudelft.nl}
	\cortext[cor1]{Senior authors contributed equally: author order decided by a coin toss. Correspondence:}

	\affiliation[inst1]{organization={Department of BioMechanical Engineering, Faculty of Mechanical Engineering, Delft University of Technology},
	}
	\affiliation[inst2]{organization={Department of Material Science Engineering, Faculty of Mechanical Engineering, Delft University of Technology},
	}

	\begin{abstract}
		Constitutive evaluations often dominate the computational cost of finite element (FE) simulations whenever material models are complex.
		Neural constitutive models (NCMs)\RV{, i.e., neural network-based constitutive models,} offer a highly expressive and flexible framework for modeling complex material behavior in solid mechanics.
		However, their practical adoption in large-scale FE simulations remains limited due to significant computational costs, especially in repeatedly evaluating stress and stiffness.
		NCMs thus represent an extreme case: their large computational graphs make stress and stiffness evaluations prohibitively expensive, restricting their use to small-scale problems.
		In this work, we introduce COMMET, an open-source FE framework whose architecture has been redesigned from the ground up to accelerate high-cost constitutive updates.
		Our framework features a novel assembly algorithm that supports batched and vectorized constitutive evaluations, compute-graph-optimized derivatives that replace automatic differentiation, and distributed-memory parallelism via MPI.
		These advances dramatically reduce runtime, with speed-ups exceeding three orders of magnitude relative to traditional non-vectorized automatic differentiation-based implementations.
		While we demonstrate these gains primarily for NCMs, the same principles apply broadly wherever for-loop based assembly or constitutive updates limit performance, establishing a new standard for large-scale, high-fidelity simulations in computational mechanics.
	\end{abstract}

	\begin{keyword}
		finite element method\sep
		batch-vectorization\sep
		neural constitutive models\sep
		high-performance computing\sep
		compute graph optimization\sep
		automatic differentiation\sep
		distributed-memory parallelism (MPI)
	\end{keyword}
\end{frontmatter}

\section{Introduction}
\label{sec:intro}

The use of neural constitutive models (NCMs), i.e., neural network-based constitutive models in solid mechanics, has gained significant traction due to their exceptional expressivity, especially when compared to traditional constitutive models.
This growing interest is largely motivated by the universal approximation theorem \cite{Cybenko1989_ee67} which states that even relatively simple neural networks can approximate arbitrary continuous functions.
This insight enables a paradigm shift in material modeling from human postulation of constitutive models to data-driven learning of material responses.

Traditionally, constitutive models were formulated by collecting limited experimental data and subsequently postulating physically admissible equations to fit this data.
This process is inherently suboptimal, as it relies on the intuition of individual mechanicians to derive suitable equations -- an approach unlikely to consistently yield the best representations of material behavior.
NCMs offer an attractive alternative: they can flexibly learn to reproduce observed material responses, obviating the need to craft distinct models for different materials manually.

Nonetheless, ensuring physical admissibility in NCMs remains crucial \RV{for model generalizability}.
\RV{This is of particular importance for the integration of NCMs in finite element (FE) solvers where non-physical material behavior typically leads to instabilities.}
Consequently, recent work has focused on incorporating mathematical constraints into neural network architectures to enforce physical principles \RV{as an inductive bias} while preserving the networks’ expressive capacity \cite{Linden2023_590d,Geuken2025,Klein2025_217d,Peirlinck2024_f38b,Thakolkaran2025_6b61,Linka2021_1735,Linka2023_6533,Klein2022_3243,Weber2021_e71e,Fuhg2024_4f5d,Tepole2025_6009,Fuhg2024_7694,Yang2025_35c4,Upadhyay2024_405d,Flaschel2023_564d,Flaschel2021_4e47,Flaschel2025_15d3,Flaschel2022_7c13,Dekhovich2023_6d32}.
Moreover, notable work has been done to obtain frameworks for appropriately training such highly-parameterized NCMs in the context of solid mechanics \cite{Flaschel2021_4e47,Thakolkaran2022_37f6,Marino2023_5b43,Flaschel2023_564d,Li2022_19a1,Meng2025_1d5e}.
As a result, NCMs have been successfully applied to model a broad range of material behaviors, including (anisotropic) hyperelasticity \cite{Thakolkaran2022_37f6,Thakolkaran2025_6b61,Klein2022_3243,Tac2022_a940,As_ad2022_6b13,Vlassis2020_4224}, viscoelasticity \cite{Rosenkranz2024_223c,Holthusen2024_2e15}, plasticity \cite{Mozaffar2019_530c,Weber2023_2c29,Xu2025_4533,Vlassis2021_7f8b}, generalized standard materials \cite{Flaschel2025_15d3}, metamaterials \cite{Zheng2024_124a}, electroelasticity \cite{Klein2024_4b6f}, and thermoelasticity \cite{Fuhg2024_55b0,Zlatic2023_7a89}.

While NCMs have demonstrated remarkable flexibility and expressivity in replicating complex material behavior, their widespread adoption is hindered by the high computational cost incurred during integration into numerical solvers, especially FE programs.
Unlike traditional constitutive models, NCMs require evaluating large computational graphs, making the calculation of stress and stiffness tensors significantly more expensive in terms of floating point operations \cite{Peirlinck2024_55e5}, thereby causing the cost of the assembly process to overshadow that of linear solves.
For example, in a standard FE calculation, comparing a Mooney-Rivlin model with an NCM trained to replicate it showed that evaluating
stress and stiffness accounted for about 5\% of the total computed time with the Mooney Rivlin model but rose to 54\% with the NCM \cite{Franke2023_1e3a}.
Even for elastic material models that require no updating of state variables, computing material behavior with an NCM can become the dominant computational bottleneck.

Without targeted improvements to how solvers evaluate NCM computations, the practical utility of these models will remain limited to small-scale offline analyses.
Bridging this gap between model fidelity and computational performance is therefore critical to enabling the routine use of NCMs in large-scale simulations across engineering and scientific domains.

While many studies have focused on improving FE solver performance through better CPU cache utilization and single instruction multiple data (SIMD) based vectorization \cite{Sun2020_5bfe,Kronbichler2012_701d,Kronbichler2019_5f0a,Castelli2021_15ba,Ljungkvist_4aa2,Munch2021_72a9}, these efforts do not address the cost of complex constitutive evaluations, assuming relatively inexpensive material models (which is not the case for NCMs).
Nonetheless, the techniques developed in those works -- particularly with regard to batched computations and vectorization -- can inspire performance optimizations for solvers utilizing NCMs.
On the other hand, several studies have employed vectorized constitutive updates for FE solves and inverse problems in solid mechanics \cite{Thakolkaran2022_37f6,Thakolkaran2025_6b61,Zheng2024_124a,Klein2024_4b6f,Xue2023_65a0,Ferreira2025_5e0c}.
However, these efforts have primarily been implemented in non-performant languages such as Python -- due to the prevalence of Python-based NCM training environments -- or using JAX \cite{jax2018github}, which lacks sufficient support for scaling across multiple compute nodes.
Moreover, they have not undergone a quantitative and systematic investigation into their scaling behavior or comprehensive performance analysis.
Lastly, some efforts \cite{Peirlinck2024_55e5,Zlatic2023_7a89,Linka2021_1735,Tac2022_a940,Rosenbusch2025_731d} have focused on retrofitting NCMs into legacy FE solvers, including commercial software such as Abaqus \cite{Dassault_Abaqus_2025} and Ansys \cite{Ansys_2025R1}.
While these solvers support user-defined material models and routines, their underlying architectures do not support vectorization strategies across multiple quadrature points and elements and therefore, suffer from computational bottlenecks.
To fully leverage the representational power of NCMs in practical applications, it is essential that the surrounding FE solver architecture be re-imagined from the ground up.

To address these challenges, we introduce a \underline{novel FE assembly algorithm} that enables \underline{batched} and \underline{vectorized} evaluation of the constitutive model.
This vectorization strategy requires simultaneous access to state variables of multiple material points, which in turn necessitates a redesign of the conventional FE element-level assembly process.
To allow fine-grained control over the performance trade-offs introduced by batching, the solver architecture includes a mechanism to explicitly manage the \textit{batch size} -- the number of constitutive updates processed in a single vectorized operation.
Typically for NCMs, the stress and stiffness are calculated by automatic differentiation, which can be slow for NCMs with large computational graphs.
Therefore, we further improve FE performance by \underline{replacing automatic differentiation (AD)} with \underline{compute graph optimized (CGO)} implementations of NCMs.
In CGO, we demonstrate that a carefully designed analytical treatment of NCMs can outperform AD by enabling efficient analytical computation of first and second derivatives.
This approach significantly reduces both memory usage and computation time.
Finally, we show that batch-vectorization is compatible with \ul{distributed-memory parallelism using message-passing interface (MPI)} \cite{mpi50}, effectively marrying SIMD-based parallelism within each compute node with MPI-based parallelism across multiple compute nodes.
This hierarchical approach to parallelization enables highly scalable and efficient large-scale simulations.
\RV{For the scope of this work, we focus on hyperelastic material behavior, while noting that the overall framework is agnostic to NCM architecture and can also be applied similarly to path-dependent NCMs.}

As a companion to this work, we introduce \textbf{COMMET} (\textit{\underline{CO}mputational \underline{M}echanics and \underline{M}achine learning \underline{T}oolbox}) -- an  open-source software incorporating the technologies introduced in this work.
Specifically, COMMET provides dedicated Python-based modules for implementing and training NCMs, as well as a C++-based FE solver \RV{which has been developed on top of the Deal.II library \cite{Arndt2023_17bf}} with batch-vectorization, CGO, and MPI parallelism for scalable simulations using NCMs.
We invite the research community to utilize and contribute to COMMET to enable broader adoption and further exploration of data-driven constitutive modeling in solid mechanics.

\RV{The remaineder of this contribution is organized as follows.
	Key background on finite elements and neural constitutive models is summarized in \sect~\ref{sec:back-and-prelim}.
	We provide details on the global and batch-vectorized assembly algorithms as well as CGO and MPI parallelization in \sect~\ref{sec:vec-fe}.
	The results of extensive benchmarks of the vectorization algorithms and developed solver are presented and discussed in \sect~\ref{sec:res}.
	In \sect~\ref{sec:demo}, we present demonstration of the developed solver's capabilities by simulating the passive filling of a human heart while using an NCM to model the material behavior.
	Finally, we provide concluding remarks in \sect~\ref{sec:con}.
}

\section{Background and preliminaries}
\label{sec:back-and-prelim}
Before detailing the batch-vectorized FE assembly algorithm for efficient NCM-based large-scale modeling (see \sect\,\ref{sec:vec-fe}), we briefly outline key preliminary knowledge on FE and NCMs in \sects\,\ref{sec:back-fe} and \ref{sec:back-nncm}, respectively.

\subsection{Finite elements for solid mechanics problems}
\label{sec:back-fe}
Here, we concisely introduce the mathematical components of FE
that are relevant to this work.
For a more complete introduction to FE, readers are referred to standard textbooks, e.g., \cite{Belytschko2014_9c06,Bonet1997_c52c,Wriggers2008_7d10}.
In the context of boundary value problems in nonlinear solid mechanics, the global displacement vector $\d$ is typically obtained by using the Newton-Raphson (NR) method (or similar gradient-based methods) to solve the weak form of the linear momentum balance discretized over a domain.
In other words, $\d$ is iteratively updated by $\d \gets \d + \dd$, where
\begin{equation}
	\K\dd = -\r\,.
\end{equation}
Here, $\K$ and $\r$ represent the global stiffness matrix and global residual vector, respectively, of the form
\begin{align}
	\K & = \begin{bmatrix}
		       K^{11}_{11} & K^{11}_{12} & K^{11}_{13} & K^{12}_{11} & K^{12}_{12} & K^{12}_{13} & \hdots \\
		       K^{11}_{21} & K^{11}_{22} & K^{11}_{23} & K^{12}_{21} & K^{12}_{22} & K^{12}_{23} & \hdots \\
		       K^{11}_{31} & K^{11}_{32} & K^{11}_{33} & K^{12}_{31} & K^{12}_{32} & K^{12}_{33} & \hdots \\
		       K^{21}_{11} & K^{21}_{12} & K^{21}_{13} & K^{22}_{11} & K^{22}_{12} & K^{22}_{13} & \hdots \\
		       K^{21}_{21} & K^{21}_{22} & K^{21}_{23} & K^{22}_{21} & K^{22}_{22} & K^{22}_{23} & \hdots \\
		       K^{21}_{31} & K^{21}_{32} & K^{21}_{33} & K^{22}_{31} & K^{22}_{32} & K^{22}_{33} & \hdots \\
		       \vdots      & \vdots      & \vdots      & \vdots      & \vdots      & \vdots      & \ddots
	       \end{bmatrix}\,, &
	\r & = \begin{bmatrix}
		       r^{1}_{1} \\
		       r^{1}_{2} \\
		       r^{1}_{3} \\
		       r^{2}_{1} \\
		       r^{2}_{2} \\
		       r^{2}_{3} \\
		       \vdots
	       \end{bmatrix}\,,
\end{align}
where $K^{IJ}_{ij}$ denotes the tangent stiffness for the pair of nodes $I,J\in\{1,\,2,\, ...,\,\nNodes\}$ in the spatial directions $i,j \in\{1,2,3\}$ and $r^{I}_i$ denotes the force residual for node $I$ in direction $i$.
These entries are given by the following expressions\RV{, which result from the discretization of the weak form of the balance of equilibrium with negligible body forces in the current configuration (see, e.g. \cite{Wriggers2008_7d10} Section 4.2.3)}:
\begin{align}
	K^{IJ}_{ij} & = \intOnDomO{\grad{\shapeFnI}_{k}\brac{\delta_{ij}\tau_{kl} + \cc_{ikjl}}\grad{\shapeFnJ}_{l}} \,, \label{eq:apprx-k} \\
	r_{i}^{I}   & = \intOnDomO{\tau_{ij}\grad{\shapeFnI}_{j}} - \intOnBoundN{\shapeFnI \bar{t}_i}\,. \label{eq:apprx-r}
\end{align}
Here, $\domO$ denotes the reference material domain,
$\shapeFnI: \RR^3\rightarrow \RR$ is the shape function of node $I$,
$\grad{\shapeFnI}$ represents the gradient of $\shapeFnI$ in the spatial configuration,
$\delta$ is the Kronecker delta,
$\btau$ is the Kirchhoff stress,
$\cc$ is the spatial stiffness tensor,
$\boundN$ is the portion of the boundary to which a traction condition is prescribed, and
$\bar{\bm{t}}$ is a prescribed traction.
Additionally, the Einstein summation convention is invoked for repeated indices.

Both $\btau$ and $\cc$ are constitutive quantities that
are functions of the deformation gradient $\F = \I + \pd{\u}{\X}$ and internal state variables, where $\I$ is the identity tensor, $\u$ is the displacement field, and $\X$ represents the position in the material domain.

The integrals in \eqs\eqref{eq:apprx-k} and \eqref{eq:apprx-r} are evaluated using numerical quadrature, i.e., as the sum of the integrand evaluated at a finite number of $\nqp$ quadrature points, within a finite number of elements $\nEl$, and weighted by the \textit{volume} of the quadrature point within that element $w^{\el,\qp}$
\begin{align}
	\intOnDomO{\tau_{ij}\grad{\shapeFnI}_{j}}                                                    & \approx \sum_{\el}^{\nEl} \sum_{\qp}^{\nqp}w^{\el,\qp}\tau^{\el,\qp}_{ij}\grad{\shapeFn ^{I, \el,\qp}}_{j}\,,\label{eq:r-quad}                                                                        \\
	\intOnDomO{\grad{\shapeFnI}_{k}\brac{\delta_{ij}\tau_{kl} + \cc_{ikjl}}\grad{\shapeFnJ}_{l}} & \approx\sum_{\el}^{\nEl} \sum_{\qp}^{\nqp}w^{\el,\qp}\grad{\shapeFn^{I, \el,\qp}}_{k}\brac{\delta_{ij}\tau^{\el,\qp}_{kl} + \cc^{\el,\qp}_{ikjl}}\grad{\shapeFn^{J, \el,\qp}}_{l}\,.\label{eq:k-quad}
\end{align}
The superscript $(\cdot)^{\el,\qp}$ denotes the evaluation at quadrature point $\qp$ in element $\el$.

Evaluating the summations in \eqs \eqref{eq:r-quad} and \eqref{eq:k-quad} lies at the heart of the so-called \textit{assembly process} in FE methods.
Traditionally, the assembly is implemented using nested for-loops: iterating over each element, then over quadrature points, and finally over pairs of nodes within each element.
In this approach, the number of constitutive calculations ($\F \rightarrow \btau, \cc$) performed at each NR iteration amounts to $\nqp\times\nEl$, which is often large.
Hence, rapid computation of the constitutive map for many material points is crucial for the performant evaluation of the necessary integrals in \eqs\eqref{eq:r-quad} and \eqref{eq:k-quad}, and hence, is crucial for a fast assembly process.

For completeness, and for comparison with the new algorithms presented subsequently, the traditional algorithm for FE assembly is presented in \alg\,\ref{alg:trad-asm}.

\begin{algorithm}
	\caption{Traditional algorithm for finite element system assembly}
	\label{alg:trad-asm}
	\begin{algorithmic}[1]
		\For{$\el = 1,\dots,\nEl$} \Comment{Loop over elements}
		\For{$\qp = 1,\dots,\nqp$} \Comment{Loop over quadrature points for element}
		\State $\F \gets \I + \sum_I \u ^{I}\otimes \Grad{\shapeFn ^{I,\el,\qp}}$ \Comment{Evaluate trial $\F$ at quadrature point}
		\State $\btau,\, \cc \gets \text{constitutive\_model}\fOf{\F}$ \Comment{Evaluate stress and stiffness at quadrature point}
		\For{$I \in \{\text{Nodes on element $\el$}\}$} \Comment{Loop over nodes for element}
		\State $r_i^I \gets r_i^I + w^{\el,\qp}\tau^{\el,\qp}_{ij}\grad{\shapeFn ^{I, \el,\qp}}_{j} $ \Comment{Add contribution to residual \eq\eqref{eq:r-quad}}
		\For{$J \in \{\text{Nodes on element $\el$}\} $} \Comment{Inner loop over nodes for element}
		\State $K^{IJ}_{ij} \gets K^{IJ}_{ij} +  w^{\el,\qp}\grad{\shapeFn^{I, \el,\qp}}_{k}\brac{\delta_{ij}\tau^{\el,\qp}_{kl} + \cc^{\el,\qp}_{ikjl}}\grad{\shapeFn^{J, \el,\qp}}_{l}$ \Comment{Add stiffness contribution \eq\eqref{eq:k-quad}}
		\EndFor
		\EndFor
		\EndFor
		\EndFor
	\end{algorithmic}
\end{algorithm}

\subsection{Neural constitutive models}\label{ssec:nnconstmod}
\label{sec:back-nncm}

\begin{figure}
	\begin{center}
		\includegraphics[width=0.9\textwidth]{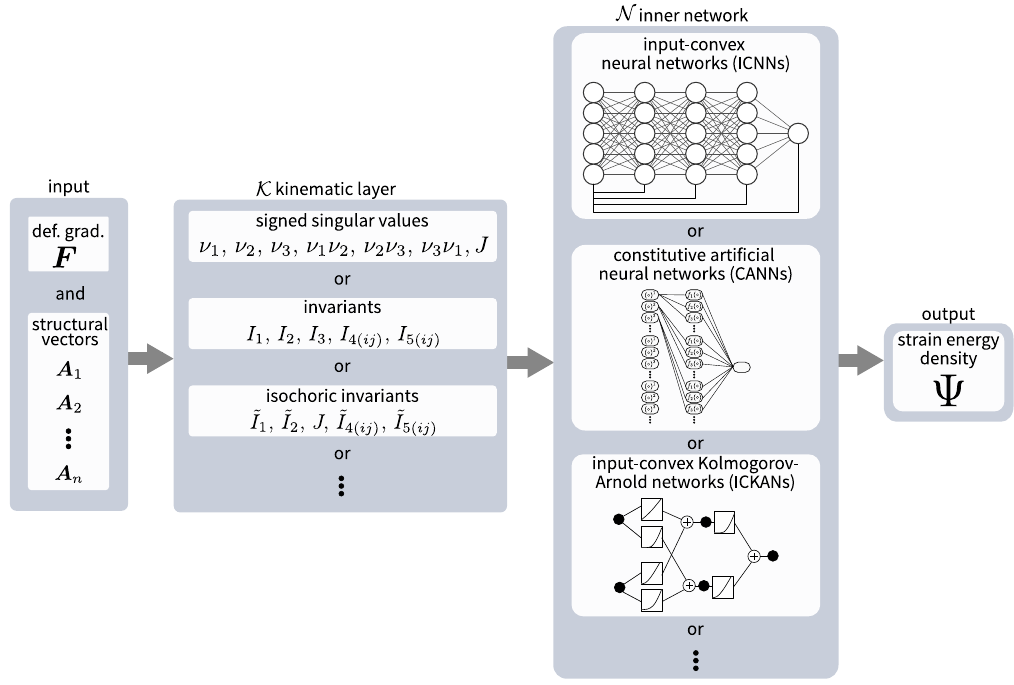}
	\end{center}
	\caption{\textbf{High-level architecture of a neural constitutive model (NCM).} The hyperelastic strain energy density formulated as a composition of two functions $\actK$ and $\actN$. The kinematic layer $\actK$ maps the deformation gradient and structural vectors to a set of invariant kinematic scalars, ensuring objectivity and material symmetry. These scalars then serve as input to the inner network $\actN$, typically a neural network architecture designed to satisfy convexity conditions required for polyconvexity. The inner network outputs the final strain energy density, which is used to derive the stress and stiffness needed in finite element simulations.}
	\label{fig:nncm-arch}
\end{figure}

For the scope of this work, we focus on hyperelastic material behavior, while noting that the overall framework is material-agnostic and can also be applied to path-dependent material behaviors.
To model (anisotropic) hyperelastic material behavior,
% \label{fn:general}
we postulate a strain energy density $\en (\F, \Aset)$ that is a function of the deformation gradient $\F$ and a set of $\nsv$ structural vectors $\Aset = \set{\StrucVec^1,\, ...,\, \StrucVec^{\nsv}}$.
The Kirchhoff stress and spatial stiffness tensor are then given by (see derivation in \ref{sec:mat-spatial})
\begin{align}
	\tau_{ij} & = \pd{\en}{F_{iJ}}F_{jJ}\,, & \cc_{ijkl} & = F_{jJ}\pdd{\en}{F_{iJ}}{F_{kL}}F_{lL} - \delta_{ik}\tau_{jl}\,.
	\label{eq:hyper-stress-stiffness}
\end{align}
In order to satisfy the axioms of objectivity and material symmetry, the strain energy density  is typically not postulated in terms of the deformation gradient $\F$ directly, but instead postulated in terms of a set of kinematic scalar values that are invariant to the choice of basis for the reference or current configuration.
% Examples of these scalars include the principal stretches $\lambda_i$, $i=1,2,3$, obtained from the spectral decomposition of $\F$ given by
% \begin{equation}
% 	\F = \sum_i \lambda_i \spaEigVec^{i} \otimes \matEigVec^{i}\,,
% \end{equation}
% where $\spaEigVec^{i}$ and $\matEigVec^{i}$ are the principal directions in the spatial and material configuration, respectively.
\RV{Examples of these scalars include the signed singular values of $\F$, $\nu_i$, $i=1,2,3$ \cite{Wiedemann2023_7b7b,Geuken2025}.\footnote{Additional requirements of $\Pi$ invariance are required for objectivity and material symmetry in this case; see \cite{Wiedemann2023_7b7b,Geuken2025} for details.}}
Alternatively, invariants of the right Cauchy-Green tensor $\C = \F^{T}\F$, e.g.,
\begin{align}
	\iI & = \tr{\C}\,, & \iII & = \frac{1}{2}\brac{\tr{\C}^{2} - \tr{\C^{2}}}\,, & \iIII & = \det{\C}\,, & \iIVind & = \StrucVeci\cdot \C \StrucVecj  \,, \label{eq:standard-inv-I}
\end{align}
where $i,j\in\{1,\dots,\nsv\}$, can also be used as inputs for the strain energy density.
Hence, in an abstract sense, the strain energy density can be thought of as a composition of two functions (see \fig\ref{fig:nncm-arch}): $\actK$ which maps the deformation gradient and structural vectors to a set of kinematic scalars, and $\actN$ which maps those scalars to the final strain energy density, i.e.,
\begin{equation}
	\en\fOf{\F, \,\Aset} =  \actNof{\actKof{\F, \,\Aset}}\,.
	\label{eq:comp-se}
\end{equation}
In the traditional constitutive modeling paradigm, $\actN$ is typically a simple analytical function that is postulated by mechanicians.
However, in the NCM paradigm, $\actN$ takes the form of a neural network architecture that is agnostic to its use as a constitutive model.
\RV{In this sense, $\actN$ is simply a highly expressive ansatz for a constitutive equation that is inspired by machine learning.}
In accordance with machine learning (ML) terminology, we term $\actK$ as the \textit{kinematic layer} and $\actN$ as the \textit{inner network}.
To ensure polyconvexity of the NCM in $\F$, certain convexity conditions are required for the inner network.
These conditions are dependent on the choice of the kinematic scalars.
For example, if the set of kinematic scalar inputs is \RV{$\set{\nu_1,\, \nu_2,\, \nu_3,\, \nu_1\nu_2,\, \nu_2\nu_3,\, \nu_3\nu_1, J}$}, then input-convexity\footnote{The architecture of an input-convex neural network \cite{Amos2017_773c} is designed to ensure that the output is identically convex with respect to the inputs, regardless of the network's weights.} of $\actN$ is sufficient to guarantee polyconvexity in $\F$ \cite{Wiedemann2023_7b7b,Tepole2025_6009,Geuken2025}.
However, if the scalar inputs are themselves polyconvex functions in $\F$ (e.g., $I_1$, $I_2$, and $I_3$), then one requires that $\actN$ is convex and monotonically increasing in its inputs \cite{Thakolkaran2025_6b61} for the NCM to be polyconvex in $\F$ overall.
To this end, some inner networks that have been used to date include input convex neural networks (ICNNs) \cite{Klein2022_3243,Thakolkaran2022_37f6,Amos2017_773c}, monotonically non-decreasing input convext neural networks (MICNNs) \cite{Jadoon2025_48c6},  constitutive artificial neural networks (CANNs) \cite{Linka2023_6533,Peirlinck2024_f38b,Peirlinck2024_55e5,Peirlinck2024_34b9}, and input convex Kolmogorov-Arnold networks (ICKANs) \cite{Thakolkaran2025_6b61,Liu2025}.
Further details on these specific NCM-based architectures are provided in \ref{sec:inner-networks} for completeness.

\section{COMMET: vectorized and batched FE solver enabling efficient NCM implementation}
\label{sec:vec-fe}

Central to our approach is a novel element assembly algorithm that enables batched and vectorized evaluation of NCMs (see \sect\,\ref{sec:fe-asm-and-batch}).
This allows the solver to fully utilize the capabilities of modern CPUs and memory hierarchies, improving cache behavior.
To enable large-scale simulations, we parallelize the solver using MPI, supporting distributed-memory computation across multiple compute nodes while maintaining consistent batched execution of NCM evaluations (see \sect\,\ref{sec:intro-mpi}).
Finally, we further reduce the cost of constitutive evaluations by replacing automatic differentiation-based computations of the stress and tangent stiffness with compute graph optimization (see \sect\,\ref{sec:intro-cgo}).

\subsection{Assembly vectorization and batching}
\label{sec:fe-asm-and-batch}

\begin{figure}
	\begin{center}
		\includegraphics[width=0.89\textwidth]{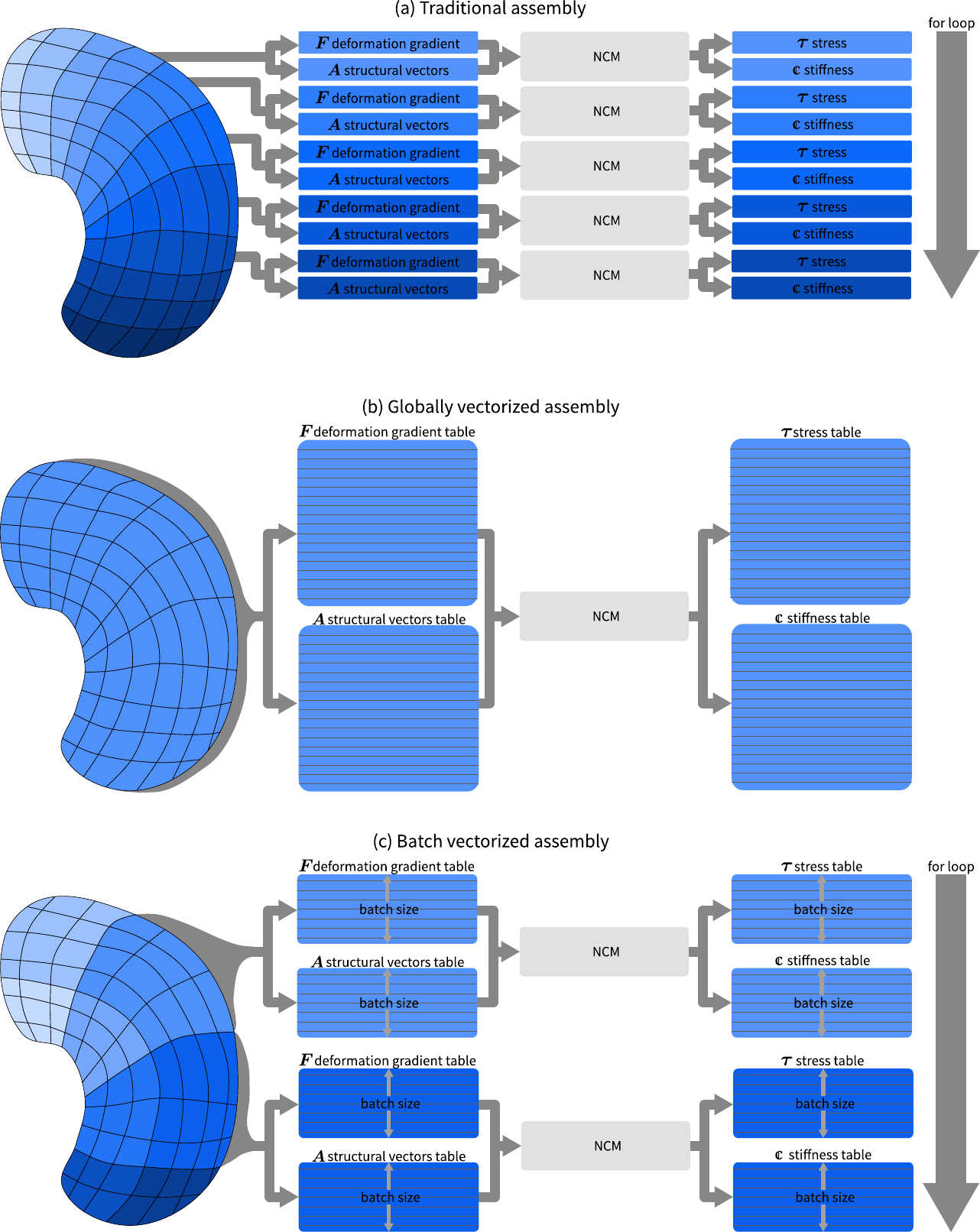}
	\end{center}
	\caption{\textbf{Schematic comparison of constitutive update strategies in finite element assembly:} (a) the traditional approach whereby the stress and stiffness are calculated for one quadrature point at a time, (b) the globally vectorized approach where the state variables (i.e. deformation gradient and structural vectors in the case of hyperelasticity) for all quadrature points are collected in tables from which associated stress and stiffness tables are calculated in a single vectorized computation, and (c) the batch-vectorized approach where batches of quadrature points are processed at a time.}
	\label{fig:batching-arch}
\end{figure}

In the traditional assembly algorithm (see \alg~\ref{alg:trad-asm}), one loops over each element and quadrature point and evaluates the constitutive updates sequentially, as shown in \fig~\ref{fig:batching-arch}~(a).
In contrast, we propose to bundle, i.e., \textit{batch} the constitutive update calculations across multiple quadrature points -- both within and across multiple elements -- and evaluating them in parallel  through a single NCM constitutive update instance (i.e., \textit{vectorization}).

The traditional FE assembly procedure (\alg~\ref{alg:trad-asm}) does not readily allow vectorization of the constitutive updates since the state variables are overwritten from one quadrature point to the next throughout the assembly procedure.
Hence, we alter this procedure by creating tables for the relevant state variables across all elements and quadrature points.
Each table is stored in a structured and contiguous block in the memory.
We introduce this assembly process as the \textbf{globally vectorized} algorithm schematically shown in \fig~\ref{fig:batching-arch}~(b).
In this algorithm, we perform the constitutive update across the entire mesh in a single vectorized operation as presented in \alg~\ref{alg:vec-asm}.

While the globally vectorized algorithm is arguably the simplest way to leverage vectorization, memory constraints only render it practical for small mesh sizes.
The RAM on a compute node is inherently limited, therefore a contiguous block of memory may not be available to store a large table of  state variables for a high-resolution mesh.
We address this issue by dividing the state variables table into multiple batches of smaller but equal sizes, with each batch still stored in a contiguous memory block.
\fig~\ref{fig:batching-arch}~(c) introduces the second \textbf{batch-vectorized} algorithm which processes the constitutive update in user-chosen batch sizes as detailed in \alg~\ref{alg:bat-vec-asm}.
Practically, we recommend using the batch-vectorized algorithm and determining the optimal batch size subject to RAM usage constraints for each machine through computational experiments, as demonstrated in \sect~\ref{sec:res-qp}.
Note that the globally vectorized algorithm is equivalent to the batch-vectorized algorithm with batch size equal to the total number of material/quadrature points (subject to memory constraints).

\begin{algorithm}[!tb]
	\caption{Algorithm for \textbf{globally vectorized} finite element system assembly}
	\label{alg:vec-asm}
	\begin{algorithmic}[1]
		\State $\text{count} \gets 0$
		\For{$\el = 1,\dots,\nEl$} \Comment{Loop over elements}
		\For{$\qp = 1,\dots,\nqp$} \Comment{Loop over quadrature points}
		\State $\vF\brac{\text{count}} \gets \I + \sum_I \u ^{I}\otimes \Grad{\shapeFn ^{I,\el,\qp}}$ \Comment{Evaluate trial $\F$ at quadrature point}
		\State $\text{count}++$
		\EndFor
		\EndFor
		\State $\ven,\,\vtau,\,\vcc \gets \text{vectorized\_constitutive\_model}\fOf{\vF}$
		\State $\text{count} \gets 0$
		\For{$\el = 1,\dots,\nEl$} \Comment{Loop over elements}
		\For{$\qp = 1,\dots,\nqp$} \Comment{Loop over quadrature points}
		\State $\btau,\, \cc \gets  \vtau\brac{\text{count}}, \vcc\brac{\text{count}}$ \Comment{Look up quadrature point values}
		\For{$I \in \{\text{Nodes on element $\el$}\}$} \Comment{Loop over nodes for element}
		\State $r_i^I \gets r_i^I + w^{\el,\qp}\tau^{\el,\qp}_{ij}\grad{\shapeFn ^{I, \el,\qp}}_{j} $\Comment{Add contribution to residual \eq\eqref{eq:r-quad}}
		\For{$J \in \{\text{Nodes on element $\el$}\} $} \Comment{Inner loop over nodes for element}
		\State $K^{IJ}_{ij} \gets K^{IJ}_{ij} +  w^{\el,\qp}\grad{\shapeFn^{I, \el,\qp}}_{k}\brac{\delta_{ij}\tau^{\el,\qp}_{kl} + \cc^{\el,\qp}_{ikjl}}\grad{\shapeFn^{J, \el,\qp}}_{l}$ \Comment{Add stiffness contribution \eq\eqref{eq:k-quad}}
		\EndFor
		\EndFor
		\EndFor
		\EndFor
	\end{algorithmic}
\end{algorithm}

\begin{algorithm}[!tb]
	\caption{Algorithm for \textbf{batch-vectorized} assembly}
	\label{alg:bat-vec-asm}
	\begin{algorithmic}[1]
		\State $\text{accumulated\_count},\,\text{count} \gets 0$
		\While{$\text{accumulated\_count}<\nEl$}
		\State $\text{count} \gets 0$
		\While{$\text{count}<\nBatch \And \text{count}+\text{accumulated\_count} <\nEl$} \Comment{Loop over element batch}
		\State $\vF\brac{\text{count}} \gets \I + \sum_I \u ^{I}\otimes \Grad{\shapeFn ^{I,\el,\qp}}$ \Comment{Evaluate trial $\F$ at quadrature point}
		\State $\text{count}++$
		\EndWhile
		\State $\ven,\,\vtau,\,\vcc \gets \text{vectorized\_constitutive\_model}\fOf{\vF}$
		\State $\text{count} \gets 0$
		\While{$\text{count}<\nBatch \And \text{count}+\text{accumulated\_count} <\nEl$}\Comment{Loop over element batch}
		\For{$\qp = 1,\dots,\nqp$} \Comment{Loop over quadrature points}
		\State $\btau,\, \cc \gets  \vtau\brac{\text{count}}, \vcc\brac{\text{count}}$ \Comment{Look up quadrature point values}
		\For{$I \in \{\text{Nodes on element $\el$}\}$} \Comment{Loop over nodes for element}
		\State $r_i^I \gets r_i^I + w^{\el,\qp}\tau^{\el,\qp}_{ij}\grad{\shapeFn ^{I, \el,\qp}}_{j} $\Comment{Add contribution to residual \eq\eqref{eq:r-quad}}
		\For{$J \in \{\text{Nodes on element $\el$}\} $} \Comment{Inner loop over nodes for element}
		\State $K^{IJ}_{ij} \gets K^{IJ}_{ij} +  w^{\el,\qp}\grad{\shapeFn^{I, \el,\qp}}_{k}\brac{\delta_{ij}\tau^{\el,\qp}_{kl} + \cc^{\el,\qp}_{ikjl}}\grad{\shapeFn^{J, \el,\qp}}_{l}$ \Comment{Add stiffness contribution \eq\eqref{eq:k-quad}}
		\EndFor
		\EndFor
		\State $\text{count}++$
		\EndFor
		\EndWhile
		\State $\text{accumulated\_count} \gets \text{accumulated\_count}+\text{count}$
		\EndWhile
	\end{algorithmic}
\end{algorithm}

\clearpage

Overall, the batch-vectorization approach leverages the Single Instruction Multiple Data (SIMD) \cite{Hennessy2011_61e5} paradigm to accelerate the FE assembly.
SIMD refers to data-level parallelism where the same instructions are executed simultaneously on multiple data points, given that there are no interdependencies between data points or their corresponding instructions.
As a simple example, the addition of two arrays can be performed in two ways: using a non-SIMD approach, where each element is processed sequentially in a loop, e.g.,
\begin{lstlisting}[language=C++]
for(unsigned int i = 0; i < array_size; i++)
    a[i] = b[i] + c[i];
\end{lstlisting}
or using the faster SIMD approach, where a single instruction adds multiple elements in parallel, e.g.,
\begin{lstlisting}[language=C++]
a[0:array_size] = b[0:array_size] + c[0:array_size];
\end{lstlisting}
Within the FE context and the proposed algorithm, the computational benefit of this SIMD approach emerges from multiple aspects, the most important ones being as follows.

\textbf{Data prefetching}
\begin{adjustwidth}{12pt}{}
	As illustrated in \fig\,\ref{fig:mem-hierarchy}, modern computers typically employ two types of RAM: dynamic random access memory (DRAM) and static random access memory (SRAM) \cite{Drepper2007_7620}.
	In general, accessing data from SRAM is roughly two orders of magnitude faster than reading data from DRAM.
	However, SRAM requires more physical space per byte and is significantly more expensive to manufacture.
	As a result, modern systems use a relatively small amount of SRAM -- ranging from a few kilobytes to several megabytes -- integrated directly into the CPU as a \textit{cache}.
	In contrast, DRAM is used for \textit{main memory}, typically on the order of gigabytes, and is connected to the CPU via a memory controller and memory bus located on the motherboard.

	The cache itself is divided into three levels, L1, L2, and L3 illustrated in \fig\,\ref{fig:mem-hierarchy}.
	These L1, L2, and L3 SRAM caches have increasing sizes and decreasing speeds.
	For example, a 24-core Intel(R) Xeon(R) Gold 6248R chip has L1, L2, and L3 has cache level sizes of
	3 MiB, 48 MiB, and 71.5 MiB, respectively, and data-retrieval latencies amounting to 4 cycles, 12 cycles, and 44 cycles, respectively \cite{SkylakeProcessors_6bba,IntelSkylake_3f49}.
	The L1 SRAM cache is further subdivided into L1D and L1I for data and instruction caching, respectively.
	Main memory, although much larger than cache, has a latency on the order of 200 cycles \cite{SkylakeProcessors_6bba,IntelSkylake_3f49}.
	Hence, memory is arranged in a hierarchical sense, with decreasing size and increasing speed from main memory to L3, L2, and L1 cache, as illustrated by the hierarchical arrows shown in \fig\,\ref{fig:mem-hierarchy}.
	\begin{figure}[!b]
		\begin{center}
			\includegraphics[width=0.5\textwidth]{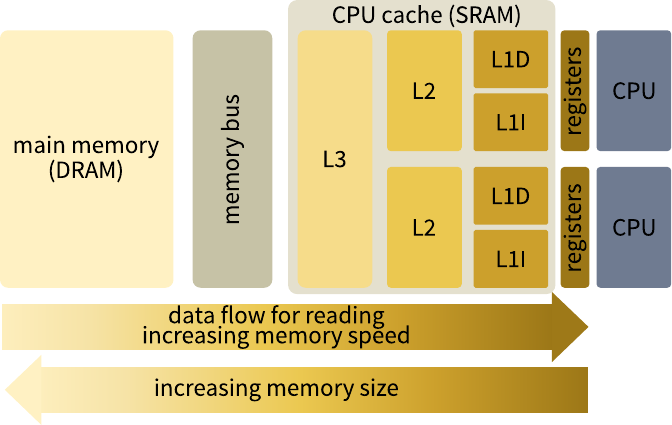}
		\end{center}
		\caption{\textbf{Schematic computer memory hierarchy with decreasing latency and increasing speed from left to right.}
			The hierarchy consists of main memory (consisting of DRAM); CPU cache (consisting of SRAM) which is further divided into L3, L2, and L1 cache; and registers.
			The L1 cache is further divided into an L1D cache for storing data and an L1I cache for storing instructions, in contrast to the L3 and L2 cache which store both data and instructions.
			Typically, each CPU has a dedicated L1 and L2 cache while the L3 cache is often shared between multiple CPUs.
		}
		\label{fig:mem-hierarchy}
	\end{figure}

	When a CPU executes an operation on data, it must first load that data into its \textit{registers}, which are extremely small and fast memory locations (typically 8 - 512 bits in size) embedded directly in the CPU.
	When loading these data into registers, the hierarchy is traversed; the L1 cache is checked first, then L2, L3, and finally, if the data is not contained in the cache, it will be retrieved from main memory.
	Obtaining the desired data from main memory as opposed to cache is termed a \textit{cache miss} and is costly as the CPU often remains idle while it waits for the necessary data to arrive.

	Modern CPUs can perform arithmetic operations far faster than they can fetch data from main memory \cite{Drepper2007_7620}.
	For instance, a floating-point addition typically completes in about one clock cycle \cite{Fog2022_4cfa}, whereas retrieving a single double-precision value from main memory can take around 200 cycles.
	This stark imbalance means that performance is often limited not by compute speed but by memory latency.
	To sustain efficiency, data must therefore be available in the CPU cache at the moment it is needed.

	To mitigate this performance gap between computation speed and memory latency, modern CPUs employ a technique known as \textit{prefetching}.
	Prefetching involves predicting which memory locations will be accessed in the near future and pre-emptively loading that data into cache before it is explicitly requested by the CPU. When data is structured contiguously in memory, such as in arrays, it enables the hardware prefetcher to recognize regular patterns and preemptively load upcoming cache lines. In contrast, if data is scattered across memory in a non-contiguous fashion -- as can occur with data structures like linked-lists or trees -- prefetching becomes much less effective, which leads to more cache misses and stalls as the CPU waits on data from main memory.

	In the context of FE assembly -- often implemented using object-oriented programming, though not exclusively -- state variables are typically passed to or stored within individual element instances. This results in a non-contiguous memory layout for the state variables. When such a layout is used with NCMs, it can lead to frequent cache misses and processor stalling due to repeated accesses to main memory. In contrast, our approach constructs a contiguous data structure for the state variables, enabling hardware prefetching to minimize cache misses and significantly improve performance.
\end{adjustwidth}

\textbf{Vector registers}
\begin{adjustwidth}{12pt}{}
	Modern CPUs are equipped with \textit{wide vector registers} that support SIMD operations.
	Common examples include 128-bit SSE2, 256-bit AVX, and 512-bit AVX-512 registers.
	Since double-precision floating-point numbers occupy 64 bits, these registers can hold 2, 4, and 8 such values, respectively.

	These CPUs also support vectorized instructions that operate on all elements within a vector register simultaneously.
	For instance, the assembly instruction \texttt{VADDPD zmm0, zmm1, zmm2} performs element-wise addition of the 512-bit registers \texttt{zmm1} and \texttt{zmm2}, storing the result in \texttt{zmm0}.
	This enables 8 double-precision additions to be performed in a single instruction.

	To use these vectorized operations, the data must be properly aligned and structured in memory to map cleanly onto the registers.
	As with prefetching, contiguous and well-aligned memory layouts are essential for maximizing the performance benefits of SIMD execution.
\end{adjustwidth}

\textbf{Overhead of launching compute kernels}
\begin{adjustwidth}{12pt}{}
	Beyond hardware-related efficiencies, software implementation choices can also introduce performance bottlenecks.
	Traditional FE methods -- without the use of neural constitutive models (NCMs) -- have typically been implemented in high-performance languages such as Fortran and C/C++.
	However, with the rise of ML in computational mechanics, NCMs are now almost exclusively implemented and trained in high-level scripting environments like Python \cite{ThePython_3e67}, where frameworks such as PyTorch \cite{Paszke2019_5564} and TensorFlow \cite{tensorflow2015_whitepaper} operate.

	To bridge this software gap for integration of NCMs in FE, recent works \cite{Zheng2024_124a,Thakolkaran2022_37f6} have developed FE and similar numerical methods around NCMs in Python-like environments.
	While the NCM component benefits from the highly optimized C/C++-like backends of these ML libraries, the remaining FE components suffer from well-known performance limitations of Python-like environments, thereby becoming computational bottlenecks.

	An underexplored alternative is exporting trained NCM models—specifically their weights and computational graphs—from Python-based ML frameworks, and importing them into high-performance languages like C/C++, where the rest of the FE code is executed efficiently.
	There are two ways that this can be done: reimplement the NCM model in the C/C++ codebase by hand, or export the NCM model via a graph-compilation tool such as TorchScript \cite{TorchScript_4549}.
    Current software environments for ML and high-performance languages like C/C++ only support the latter as hard-coding neural networks with complex architectures and \RV{large numbers} of parameters would be too labor-intensive.

    \RV{Each call that is made to the graph-compiled model requires several overhead operations, e.g. the data type, layout, and contiguity of the input tensors is checked, reference counters are updated, memory is allocated, etc. 
    The cost of these operations is known as the ``compute kernel launch overhead''.
    This cost is distinct from the cost of instantiating an NCM and loading its weights from memory, i.e. even for a single NCM instance the compute kernel launch overhead is incurred for each call made to that instance.
    In traditional FE assembly algorithms (see \alg~\ref{alg:trad-asm}), the compute kernel is launched for every material point \cite{Peirlinck2024_55e5}, and so the incurred compute kernel launch overhead is multiplied by the number of material points.}

	To mitigate this, we leverage vectorization by processing multiple material points simultaneously. 
    This approach reduces the incurred kernel launch overhead significantly, requiring only one invocation per batch, and thus achieving a speed-up proportional to the batch size.

\end{adjustwidth}

Through low-level monitoring tools and CPU performance counters, one can quantify the benefits of individual optimization aspects using micro-benchmarks.
However, in the context of FE methods, these aspects are often tightly intertwined, making it difficult to isolate their individual contributions.
As a result, our work focuses on the overall speedup achieved from all contributing factors combined, while acknowledging that a detailed breakdown of each low-level contribution is beyond the scope.

\begin{remark}
	\RV{The necessity of instructions being identical across all data poses a challenge for constitutive updates that may have control flow which differs from one material point to the next, e.g. updates that require local Newton-Raphson iterations for the so-called return mapping algorithm in plasticity.
		A possible solution in this case is to continue iterating across the whole batch until the state variables for all material points have converged.
		We note, however, that many path dependent NCMs that have been presented in the literature do not require solving of nonlinear equations for the evolution state variables through such iterative schemes \cite{Rosenkranz2024_223c,Guo2025_3df3,Mozaffar2019_530c,Weber2023_2c29}, and so the proposed global and batch-vectorized algorithms can be applied as is.
	}
\end{remark}

\subsection{Compute graph optimization}
\label{sec:intro-cgo}

When using NCMs in FE solvers, it is necessary to compute the first and second derivatives of the strain energy density function to obtain the stress and tangent stiffness in accordance with \eq\,\eqref{eq:hyper-stress-stiffness}.
While automatic differentiation (AD) is a widely adopted approach for this purpose due to its accuracy and ease of integration \cite{Ferreira2025_5e0c,Seidl2022_3da1,Al_Hassanieh2025_2f66,Chen2023_5928}, it incurs significant performance overhead — particularly in reverse-mode implementations \cite{Eval2008Ch3,Eval2008Ch10} required for computing second-order derivatives (Hessians).
This overhead stems from the need to construct and traverse computational graphs multiple times, along with the storage of intermediate states during evaluation, which together lead to increased memory consumption and compute time.
Numerical differentiation via input perturbation offers no clear advantage: it typically requires multiple evaluations of the NCM, and does not provide accurate gradients, potentially leading to unstable simulations.
Analytically derived gradients, in contrast, can significantly reduce both memory usage and computational cost.
However, unlike traditional constitutive models, they are not straightforward to obtain for NCMs with large computational graphs.

To further reduce execution time of the constitutive updates, we introduce a general framework to obtain analytical derivatives of NCMs with just one forward pass (i.e., a single evaluation of the NCM).
This is significantly faster than AD while providing exact gradients unlike numerical differentiation.
We term implementations that make use of such analytical derivations as \textit{compute graph optimization} (CGO).
Without loss of generality, this work presents the formulation for hyperelasticity, with the extension to path-dependent materials identified as a direction for future research.

The core idea is to compute the intermediate derivatives of the network layers in a modular fashion using chain rule of differentiation.
We refer back to Section \ref{ssec:nnconstmod} and \fig~\ref{fig:nncm-arch}, where the strain energy density is described as a composition of a kinematic layer $\actK$ and inner network $\actN$ \eq\eqref{eq:comp-se}.
Accordingly, we obtain the following expressions:
\begin{align}
	\pd{\en}{F_{iJ}} & = \pd{\actN}{\actK_m}\pd{\actK_m}{F_{iJ}}\,, & \pdd{\en}{F_{iJ}}{F_{kL}}= \pdd{\actN}{\actK_m}{\actK_n}\pd{\actK_m}{F_{iJ}}\pd{\actK_n}{F_{kL}} +  \pd{\actN}{\actK_m}\pdd{\actK_m}{F_{iJ}}{F_{kL}}\,.
	\label{eq:nncm-chain}
\end{align}
Hence, we can implement the first and second derivatives for different kinematic layer types (principal strains, invariants, etc.) and  inner network types (different  architectures) individually and then combine them through \eq\eqref{eq:nncm-chain} to obtain the associated computational graph for the gradients of the strain energy density, thus providing modularity.

We provide associated derivatives for various types of kinematic layers in \ref{sec:kinnematic-layers}.
As didactic illustrations, we present the analytical first and second derivatives for two notable hyperelastic NCMs found in the literature, MICNNs \cite{Klein2022_3243,Thakolkaran2022_37f6,Amos2017_773c,Jadoon2025_48c6} and CANNs \cite{Linka2023_6533,Peirlinck2024_f38b,Peirlinck2024_34b9,Peirlinck2024_55e5} in \ref{sec:inner-networks}.
These derivations can be similarly extended for other NCMs.

\subsection{Compatibility with distributed-memory parallelization for scalable simulations}
\label{sec:intro-mpi}

Parallelization is the primary strategy for accelerating scientific computing applications.
On shared-memory machines (workstations), where all CPUs have access to a common memory space, parallelization is typically achieved by employing multiple threads within a single process.
This approach is effective as long as the entire problem fits within the memory of a single compute node.

However, computational requirements for large-scale simulations, such as finite element analyses on highly refined meshes, often exceed the memory and processing capacity of a single motherboard.
In such cases, computations are performed on distributed-memory systems, or supercomputers, which consist of multiple compute nodes connected by high-speed interconnects.
Each node has its own private memory, and processes on one node cannot directly access the memory of another.
As a result, thread-based parallelization within a single process is insufficient.
Instead, parallel execution requires launching at least one process per compute node, with data exchange between processes handled explicitly.
This communication is conventionally managed through the Message Passing Interface (MPI) \cite{Balay2025_3565,Arndt2023_17bf,mofemJoss2020,BarattaEtal2023}, a standardized and portable message-passing system that has become the de facto approach for distributed scientific computing.

\begin{figure}[!tb]
	\begin{center}
		\includegraphics[height=0.63\paperheight]{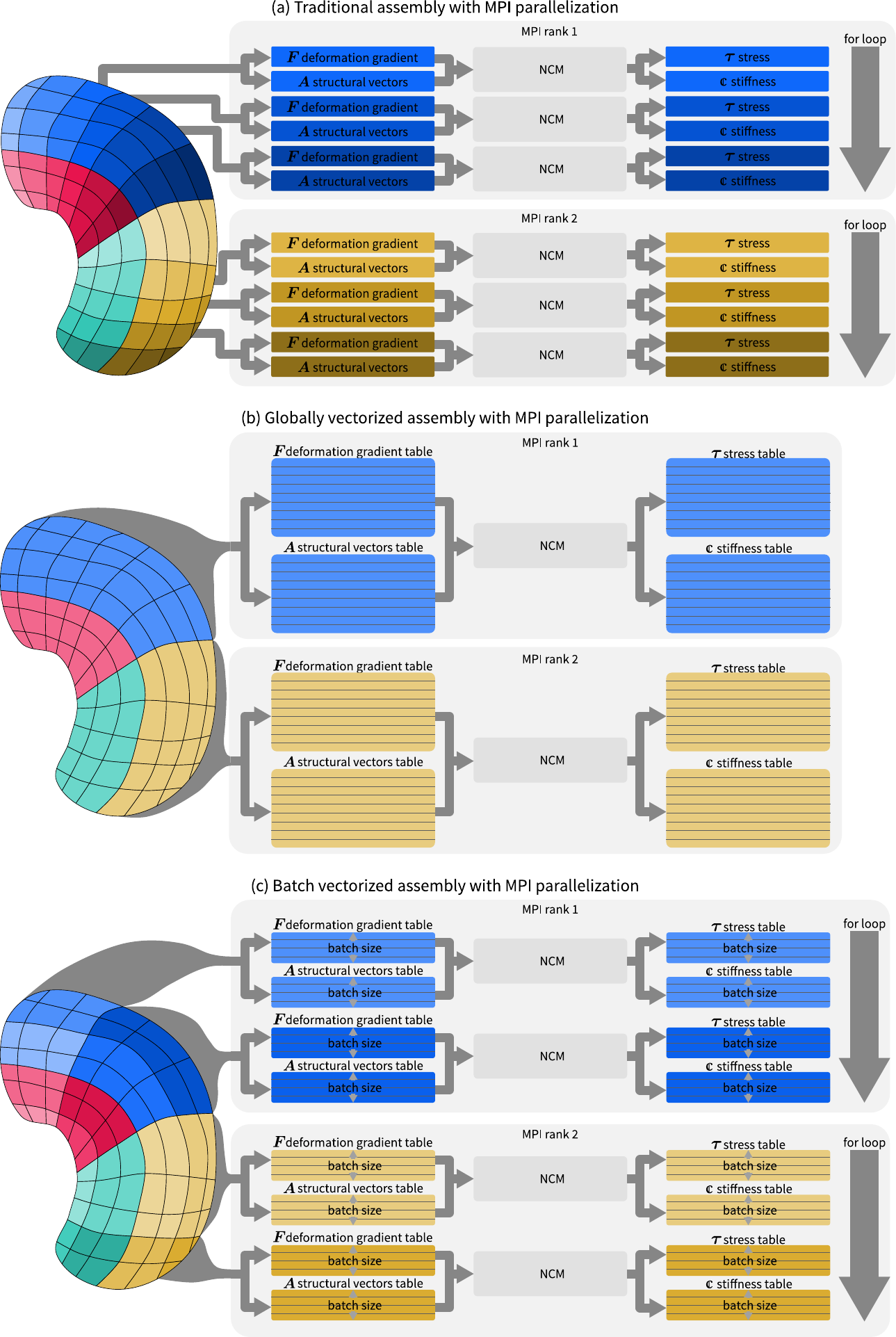}
	\end{center}
	\caption{\textbf{Schematic comparison of constitutive update procedures in finite element assembly under MPI-based distributed parallelization:} (a) traditional, (b) globally vectorized, and (c) batch-vectorized algorithms apply similarly to the single process case shown in \fig\,\ref{fig:batching-arch}.
		However, each MPI rank is only responsible for assembly on its associated subdomain of the mesh.
		Accordingly, for the globally vectorized algorithm (b), the table sizes correspond to the subdomain owned by the rank as opposed to the entire mesh.}
	\label{fig:mpi-fe}
\end{figure}

Our proposed algorithms for vectorized assembly with NCMs are fully compatible with MPI-based distributed computing.
As illustrated in \fig\,\ref{fig:mpi-fe}, the computational mesh is partitioned into subdomains, each assigned to a specific MPI rank (i.e., process).
The globally and batch-vectorized assembly algorithms are then executed independently on each rank for its local subdomain.
While this strategy increases the number of kernel launches, these launches occur concurrently across ranks and therefore add little to the overall overhead.

\newpage
\section{Results}
\label{sec:res}

For the purposes of benchmarking, we investigate three NCM architectures for which architecture and implementation details were discussed Section \ref{ssec:nnconstmod}:
MICNNs \cite{Klein2022_3243,Thakolkaran2022_37f6,Amos2017_773c,Jadoon2025_48c6}, CANNs \cite{Linka2023_6533,Peirlinck2024_f38b,Peirlinck2024_34b9,Peirlinck2024_55e5}, and ICKANs \cite{Thakolkaran2025_6b61,Liu2025}.
These architectures are each trained on several (simulated) hyperelastic material data in an \textit{unsupervised} manner -- as presented in the NN-EUCLID framework of Thakolkaran et al. \cite{Thakolkaran2022_37f6,Thakolkaran2025_6b61}.
Specifically, the training data contains full-field displacements and global reaction forces for test coupons containing heterogeneous strain fields.
The NCMs are then trained to satisfy the weak form of the linear momentum balance.
For demonstration purposes and without loss of generality, we assume the ground-truth material model to be a Gent-Thomas hyperelastic material.
Further details on the synthetic data generation and training of the NCMs are provided in ~\ref{sec:data-gen-and-training}.

We emphasize that the framework is agnostic to the specific NCM architecture and, more importantly, to the source of the training data. Moreover, the accuracy of the NCMs is not the focus of this work; prior and contemporary studies (e.g., \cite{Geuken2025,Tac2024_61c2}) compare the performance of various NCMs and explore strategies and architectural enhancements aimed at improving NCM fidelity.

To investigate the effects of vectorization, batching, and CGO, computational experiments are first run only at the material point (i.e., quadrature point) level and in the absence of a FE assembler and solver to isolate the mechanisms for any potential speed-ups.
We term these experiments \textit{material point benchmarks} (\sect\,\ref{sec:res-qp}).
We then investigate these effects within the context of a FE solver (\sect\,\ref{sec:res-fe}), followed by MPI scaling (\sect\,\ref{sec:res-mpi}) to verify that we can utilize these mechanisms to dramatically speed up the FE solver as a whole, while maintaining scalability on large-scale distributed computing.

\subsection{Material point benchmarks}
\label{sec:res-qp}

\subsubsection{Material point benchmarks: Effect of vectorization and batching, without CGO} \label{sec:res-vec-batch}

To evaluate the performance gains from vectorization and batching, we conduct computational experiments where constitutive updates are computed for a fixed number of material points, corresponding to quadrature points in the FE context.
These updates are computed in batches of varying sizes, referred to as ``batch size''.
Each constitutive update includes computing the strain energy density, stress tensor, and stiffness tensor.
In this benchmark, the stress and stiffness tensors are computed using AD, though these will later be replaced by CGO.

\begin{figure}
	\begin{center}
		\includegraphics[width=\linewidth]{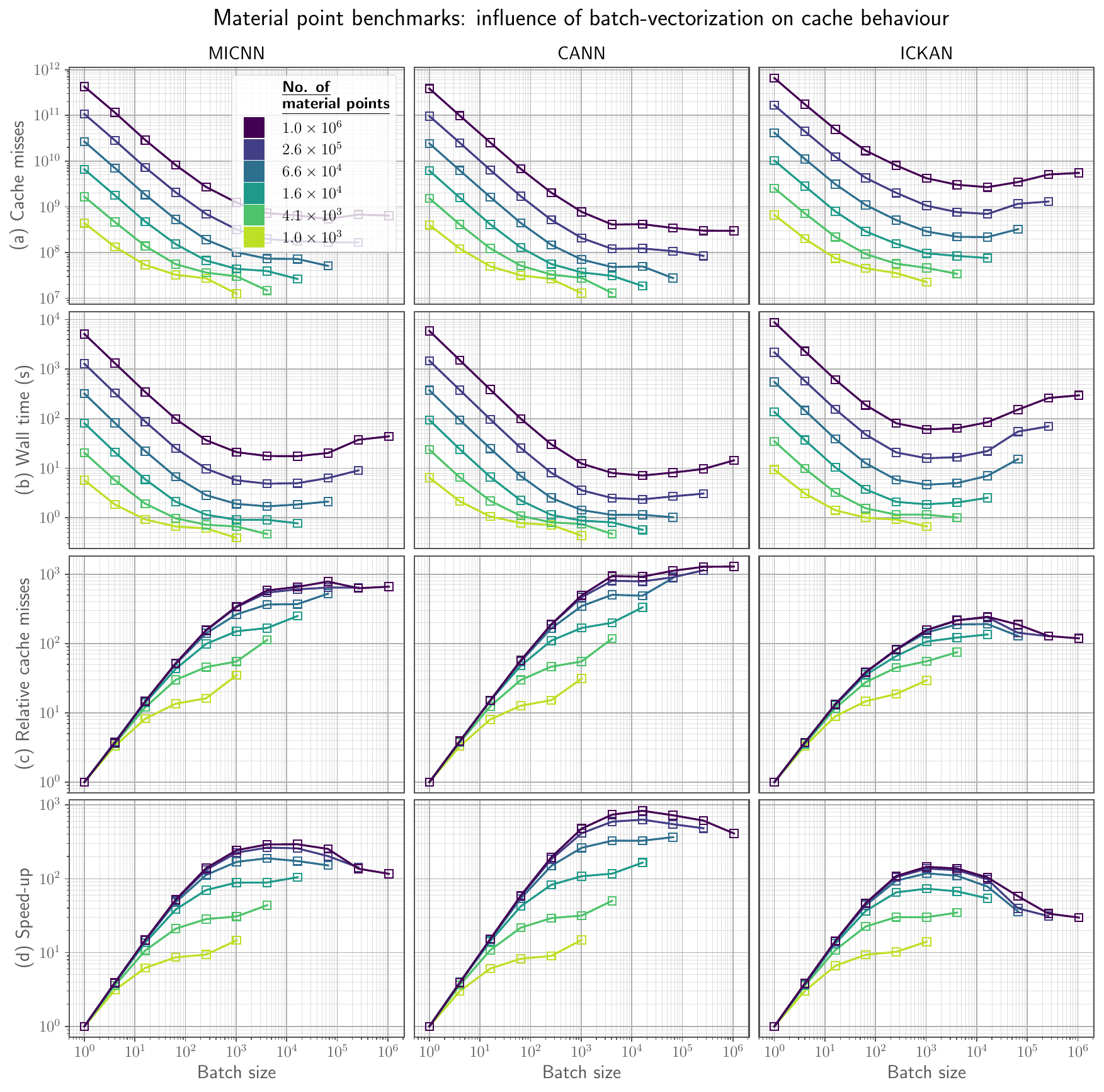}
	\end{center}
	\caption{\textbf{Effect of batch size on cache efficiency and computational performance of NCMs.} Results are shown for various fixed numbers of materials points (indicated in the legend). Metrics include (a) cache misses, (b) wall time, (c) relative cache misses (non-vectorized divided by vectorized), and (d) relative speed-up (non-vectorized wall time divided by vectorized wall time), are reported.}
	\label{fig:cache-misses}
\end{figure}

For each experiment, we report the absolute number of cache misses, cache misses relative to the non-vectorized baseline, absolute wall time, and speed-up in wall time compared to the non-vectorized baseline in \fig~\ref{fig:cache-misses}.
Here, the non-vectorized baseline refers to the sequential evaluation of constitutive updates for each material point in a loop, which is equivalent to a batch-vectorized computation with a batch size of 1.
All material point benchmarks are performed on a workstation equipped with an AMD Ryzen 9 7950X CPU, the details of which are presented in \tab~\ref{tab:deets} in \ref{sec:mach-deets}.

Increasing the batch size from $1$ to approximately $10^{3}$ reduces the number of cache misses by two orders of magnitude (\fig~\ref{fig:cache-misses}\,(a,c)).
The effect of reduction in cache misses is directly reflected in reduction of wall time and increase in \textbf{speed-up by two orders of magnitude} (\fig\,\ref{fig:cache-misses}\,(b,d)) over the non-vectorized baseline.
The similarity between reduction in cache misses and wall times---i.e. the similarity between \fig~\ref{fig:cache-misses}\,(a) and (b) and between \fig~\ref{fig:cache-misses}\,(c) and (d)---are indicative of hardware level optimizations, such as prefetching, as discussed in \sect~\ref{sec:fe-asm-and-batch}.
\ul{We emphasize that these results are obtained on a single processor, without the use of GPUs, multi-threading, or multiprocessing.}

Moreover, we observe that there is an optimal batch size around $\sim10^3$-$10^4$ for which the wall time per material point evaluation is minimized.
In the context of an FE solver, this motivates for functionality allowing for control of the batch size, so that a user may choose the optimal batch size for a given machine. Note that the optimal batch size may vary across different machines.

\begin{figure}
	\begin{center}
		\includegraphics[width=\linewidth]{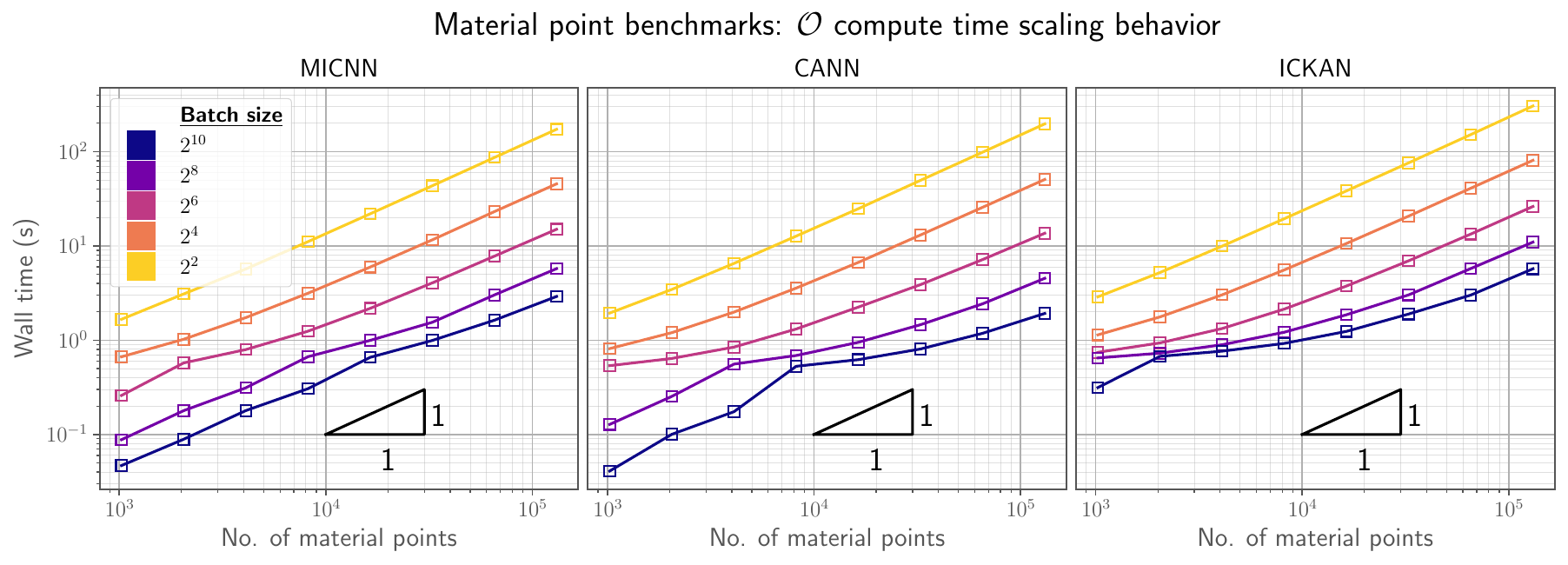}
	\end{center}
	\caption{\textbf{Scaling of compute time for various batch sizes and NCMs.} Wall time grows linearly with the number of material points. Increasing the batch size reduces the wall time but maintains the linear scaling with the number of material points.}
	\label{fig:mp-scaling}
\end{figure}

The computational complexity for non-vectorized constitutive updates is simply $\bigO\fOf{\npts}$ where $\npts$ is the number of material points.
To determine how batch size affects computational complexity, we perform computational experiments where we fix batch size and track wall time for increasing $\npts$ in \fig~\ref{fig:mp-scaling}.
The batch size reduces the wall time without affecting its linear scaling with respect to $\npts$.
Therefore, we estimate  computational complexity of batch-vectorized constitutive updates as $\bigO\fOf{m\npts}$ where $m<1$ is a factor
that depends on  the choice of  batch size and computer architecture.

We note that cache misses, speed-ups, and scaling are comparable across diverse NCM architectures---namely, MICNNs \cite{Klein2022_3243,Thakolkaran2022_37f6,Amos2017_773c,Jadoon2025_48c6}, CANNs \cite{Linka2023_6533,Peirlinck2024_f38b,Peirlinck2024_34b9,Peirlinck2024_55e5}, and ICKANs \cite{Thakolkaran2025_6b61,Liu2025}---supporting a reasonable conclusion that the observed performance gains are largely agnostic to the choice of NCM architecture.
This suggests that, for accelerating FE simulations integrated with NCMs, implementation-level optimizations---such as vectorization---can have a greater impact on inference speed than architectural enhancements like pruning or the choice of NCM architecture itself.
In light of this result and for the sake of brevity, we present the performance analyses from \sect~\ref{sec:res-pt-cgo} onward using only the MICNN architecture, while noting that similar performance gains are observed for CANNs and ICKANs.

\subsubsection{Material point benchmarks: Effect of compute graph optimization}\label{sec:res-pt-cgo}

\begin{figure}[h]
	\begin{center}
		\includegraphics[width=\linewidth]{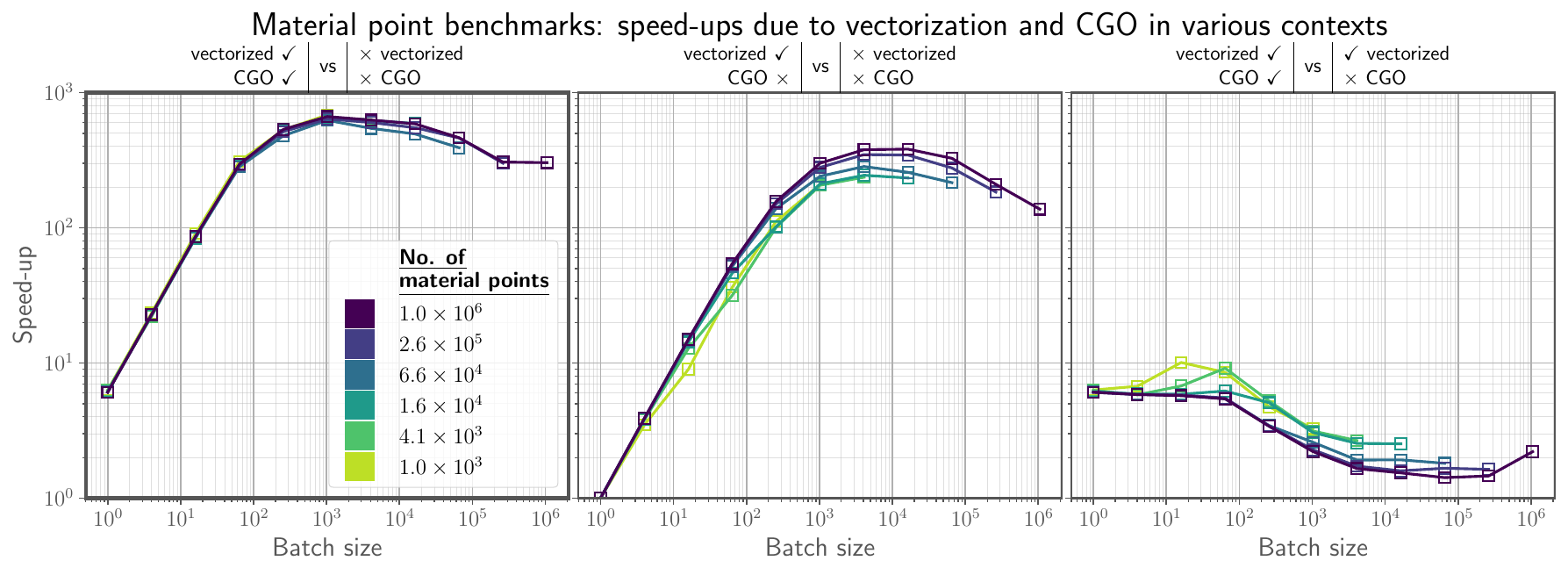}
	\end{center}
	\caption{\textbf{Performance improvements from compute graph optimization (CGO):}
		(left) combined speed-ups from CGO and vectorization relative to a non-vectorized, non-CGO baseline;
		(middle) speed-ups due to vectorization only (middle);
		and (right) Additional speed-ups due to CGO only at different batch sizes.
		Combining CGO and vectorization yields a speed-up of up to three orders of magnitude, whereas vectorization alone yields two orders of magnitude speed-up.}
	\label{fig:mp-results-w-cgo}
\end{figure}

As an alternative to AD, we use CGO (see \sect~\ref{sec:intro-cgo}) to obtain stress and stiffness tensors from a pre-trained NCM.
To demonstrate the performance improvement due to CGO, the computational experiments at the material point-level presented in \sect~\ref{sec:res-vec-batch} and \fig~\ref{fig:cache-misses} are repeated for CGO implementations and compared against AD implementations in \fig~\ref{fig:mp-results-w-cgo}. For fair comparison,  the reverse-mode AD framework is implemented using  LibTorch \cite{C_2a28}, which we assume to be highly optimized and state-of-the-art at the time of writing.

Combined batch-vectorization and CGO yields \textbf{speed-ups of nearly three orders of magnitude} over the equivalent non-vectorized non-CGO implementation (\fig~\ref{fig:mp-results-w-cgo}\,(left)), e.g. 665 times faster at batch size of $2^{10}$.
To isolate the effect of CGO alone, we benchmark the speed-up for batch-vectorized and CGO relative to batch-vectorized and non-CGO implementation (\fig~\ref{fig:mp-results-w-cgo}\,(right)) and observe speed-up between 2-10 times depending on the batch size.
The remaining speedup (up to 400 times depending on batch size) can be attributed to batch-vectorization -- as shown in benchmark of batch-vectorized and non-CGO relative to non-batch-vectorized and non-CGO implementation (\fig~\ref{fig:mp-results-w-cgo}\,(middle)).

\begin{figure}[h!]
	\begin{center}
		\includegraphics[width=\linewidth]{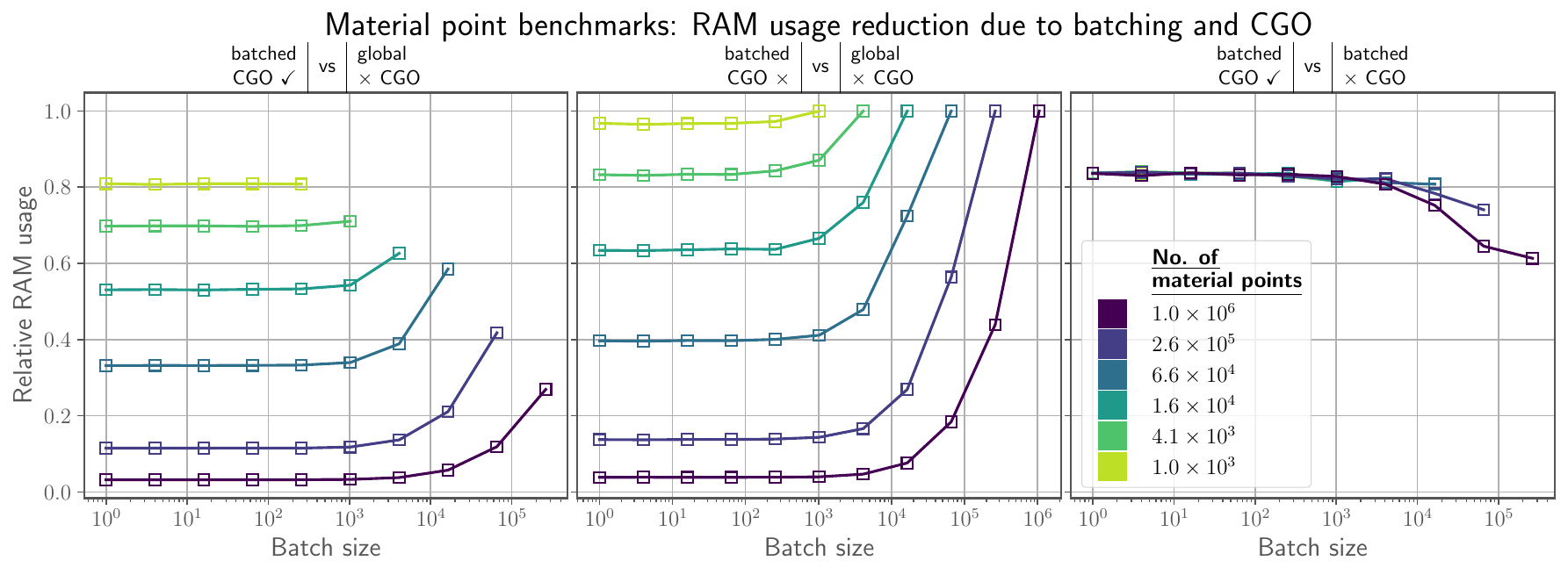}
	\end{center}
	\caption{\textbf{Reduction of RAM usage by batch-vectorization and CGO:} (left) RAM usage of batch-vectorized CGO computations relative to globally vectorized non-CGO computations; (middle) batch-vectorized non-CGO computations relative to globally vectorized non-CGO computations, and (right) Batch-vectorized CGO computations relative to batch-vectorized non-CGO computations.
		Batch-vectorization and CGO reduce RAM usage by more than 90\% at $\sim10^{6}$ material points.}
	\label{fig:mp-results-ram}
\end{figure}
In addition to reducing wall time, CGO and batching allows for a reduction in RAM usage.
To demonstrate this, we compare the RAM usage of batch-vectorized computations with and without CGO in \fig~\ref{fig:mp-results-ram}.
\textbf{The combination of batch-vectorization and CGO dramatically reduces RAM usage relative to global vectorization with non-CGO computations} (\fig~\ref{fig:mp-results-ram}\,(left)).
For example, the RAM usage is reduced by over 90\%, when using batch-vectorization and CGO in comparison to globally vectorized non-CGO implementations for $\sim 10^{6}$ material points   and  batch sizes of less than $10^{4}$ (\fig~\ref{fig:mp-results-ram}\,(left, middle)).
Additionally, we note that the optimal batch size for speed-up is in the range of $10^{3}$--$10^{4}$ (\fig~\ref{fig:mp-results-w-cgo}).
CGO alone accounts for $\sim$18--38~\% of the reduction in RAM usage (\fig\ref{fig:mp-results-ram}\,(right)).
The majority of the reduction in RAM usage is typically due to batching (\fig\ref{fig:mp-results-ram}\,(middle)), however, this largely depends on the total number of material points.

\subsection{FE benchmarks}
\label{sec:res-fe}

\begin{figure}[h]
	\center
	\includegraphics[width=.7\linewidth]{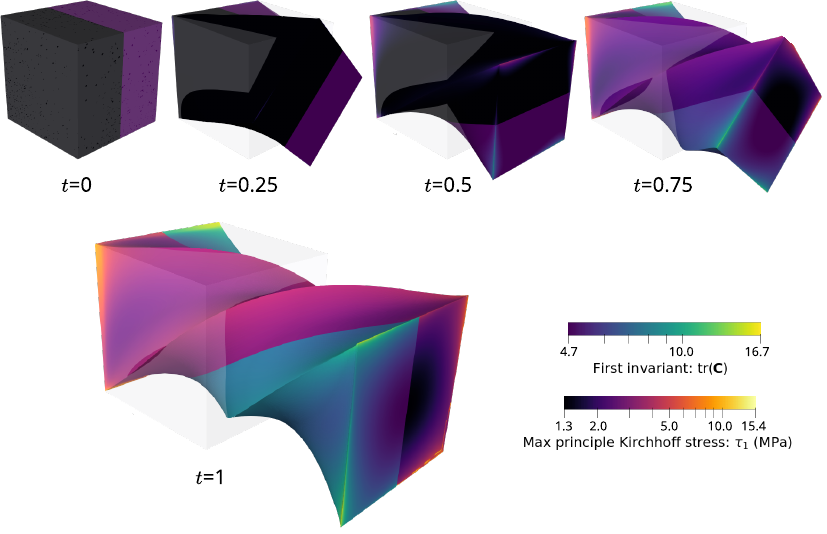}
	\caption{\textbf{Finite element benchmark problem for COMMET performance testing.} A unit cube is fixed at one end and subjected to a unit displacement and half-rotation at the other. The reference configuration is shown in grey. Large values of $\tr{\C}$ (greater than 16) demonstrate the robustness of the underlying FE solver.}
	\label{fig:twist-cube}
\end{figure}

The material point benchmarks displayed thus far elucidate the effects of vectorization, batching, and CGO on the computational cost of constitutive updates of material points when using NCMs.
We now translate this cost and speed-up in the context of FE simulations.
We propose a benchmark simulation utilizing NCMs showcased in \fig~\ref{fig:twist-cube}.
We simulate a unit cube, fixed at one end, and subjected to a unit axial tensile displacement and a half-rotation at the opposite end.
We then proceed to apply vectorization, batching, and CGO, as detailed in \sect\,\ref{sec:vec-fe}, and systematically observe the effects on the computational cost. Specifically, we compare the cost of assembly with that of solving the resulting linear system, as these are the two most computationally intensive tasks in FE simulations.

For a fair assessment, the linear solver (as part of the nonlinear Newton-Raphson solver) used in these experiments is provided by PETSc \cite{Balay2025_3565}, a highly mature, optimized, and state-of-the-art library (at the time of writing) for solving linear systems of equations.
Specifically, we use a conjugate gradient linear solver with Jacobi relaxation as the preconditioner.
	{We emphasize that the proposed framework for accelerating NCMs in finite elements is agnostic to the choice of linear solver and preconditioner; the selections here are intended merely as representative examples.}

\subsubsection{FE benchmarks: Speed-up due vectorization, batching, and CGO}\label{sec:res-fe-vec-batch}

\begin{figure}
	\begin{center}
		\includegraphics[width=.85\linewidth]{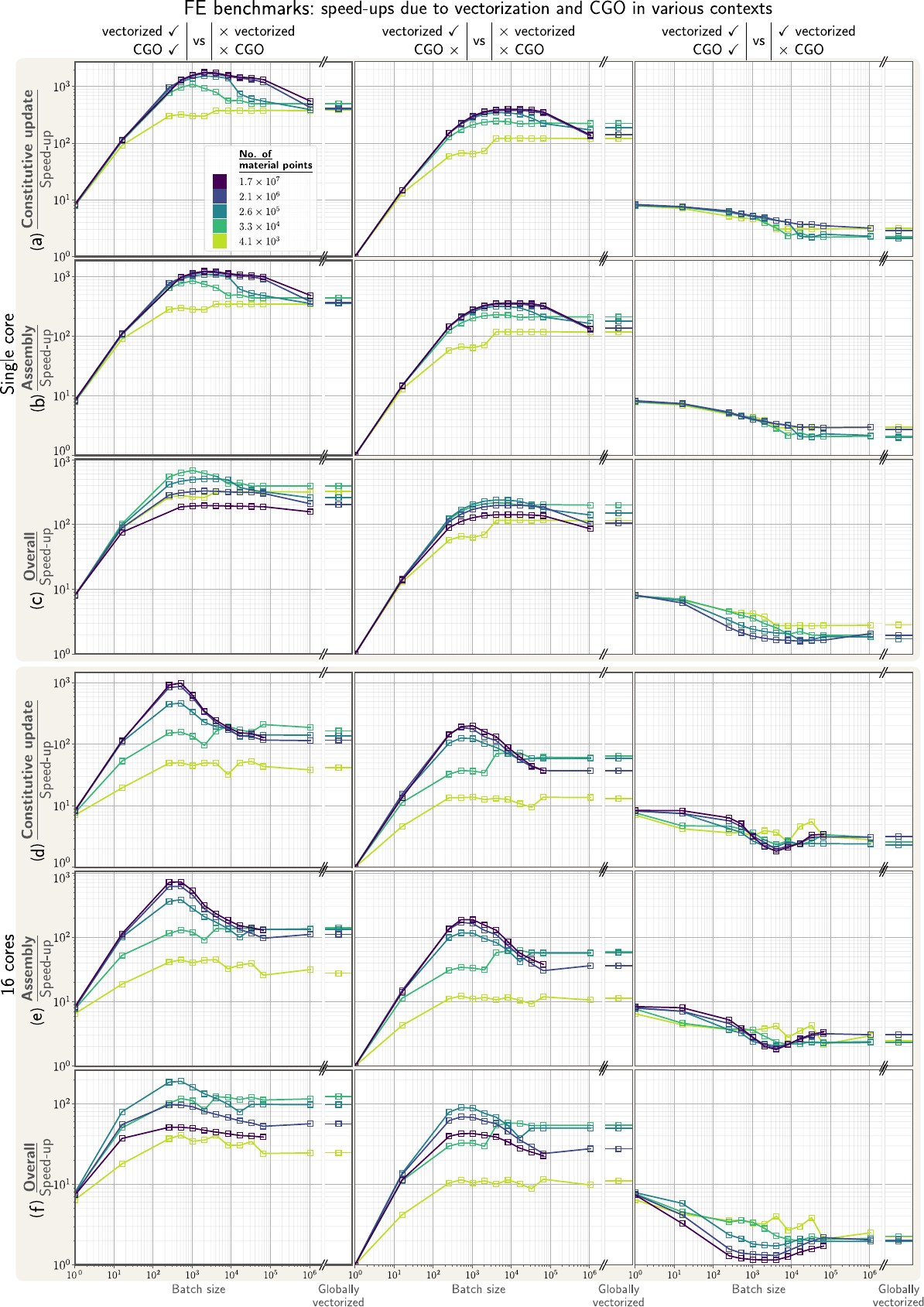}
	\end{center}
	\caption{\textbf{Speed-up of FE simulations due to vectorization and CGO:} (left column) the combined effects of vectorization and CGO, (middle column) the effects of vectorization only, and (right column) the effects of CGO only at different vectorization contexts are shown in the (top group) single core and (bottom group) multicore contexts.
		Additionally, the effects are shown for different sections of the program: (top) constitutive update, (middle) assembly, and (bottom) overall.}
	\label{fig:fe-exp-speed-up}
\end{figure}

We benchmark the computational performance of the globally-vectorized (\alg~\ref{alg:vec-asm}) and batch-vectorized (\alg~\ref{alg:bat-vec-asm}) assembly algorithms in \fig~\ref{fig:fe-exp-speed-up}, for both single-core and 16-core multiprocess (via MPI on a single node) simulations.

\textbf{Combining batch-vectorization and CGO yields a three orders of magnitude speed-up in the constitutive update} (\fig~\ref{fig:fe-exp-speed-up}\,(a), left).
The speed-ups are slightly diluted by the time taken up by other tasks in assembly (\fig~\ref{fig:fe-exp-speed-up}\,(b), left), and then diluted further in the overall simulation time (\fig~\ref{fig:fe-exp-speed-up}\,(c), left).
However, the overall speed-up is still upwards of two orders of magnitude. (\fig~\ref{fig:fe-exp-speed-up}\,(c), left).

\textbf{The majority of the speed-up, i.e. two of the three orders of magnitude, results from assembly vectorization} (\fig~\ref{fig:fe-exp-speed-up}\,(a), middle).
This is determined by repeating the computational experiments without the use of CGO and determining the speed-up relative to non-vectorized computations, thus isolating the effect of vectorization (\fig~\ref{fig:fe-exp-speed-up}\,(a), middle).
Comparing vectorized CGO computations with vectorized non-CGO computations (\fig~\ref{fig:fe-exp-speed-up}\,(a), right) isolates the speed-ups due to CGO in different vectorization contexts, and shows that CGO accounts for a speed-up of 2-10 times depending on the batch size.

\textbf{Batch-vectorization at the optimal batch size ($\sim10^{3}$--$10^{4}$) consistently outperforms global-vectorization} by a factor of 1-10  (\fig~\ref{fig:fe-exp-speed-up}\,(a-c) left).
This motivates the usage of the batch-vectorized algorithm for not only alleviating memory constraints, but also for the reduction of compute time.

\textbf{Speed-up is maintained in the multiprocessing context} (\fig~\ref{fig:fe-exp-speed-up} (d-f)).
However, the optimal batch size is shifted to be in the range of $10^{2}$--$10^{3}$ in contrast to single core case i.e. $10^{3}$--$10^{4}$.
We suspect that this results from the cores sharing the L3 cache; since multiple processes are utilizing the cache, the batch size per process must be smaller for the data to fit in the cache than in the single core/process case.

\begin{figure}
	\begin{center}
		\includegraphics[width=0.8\linewidth]{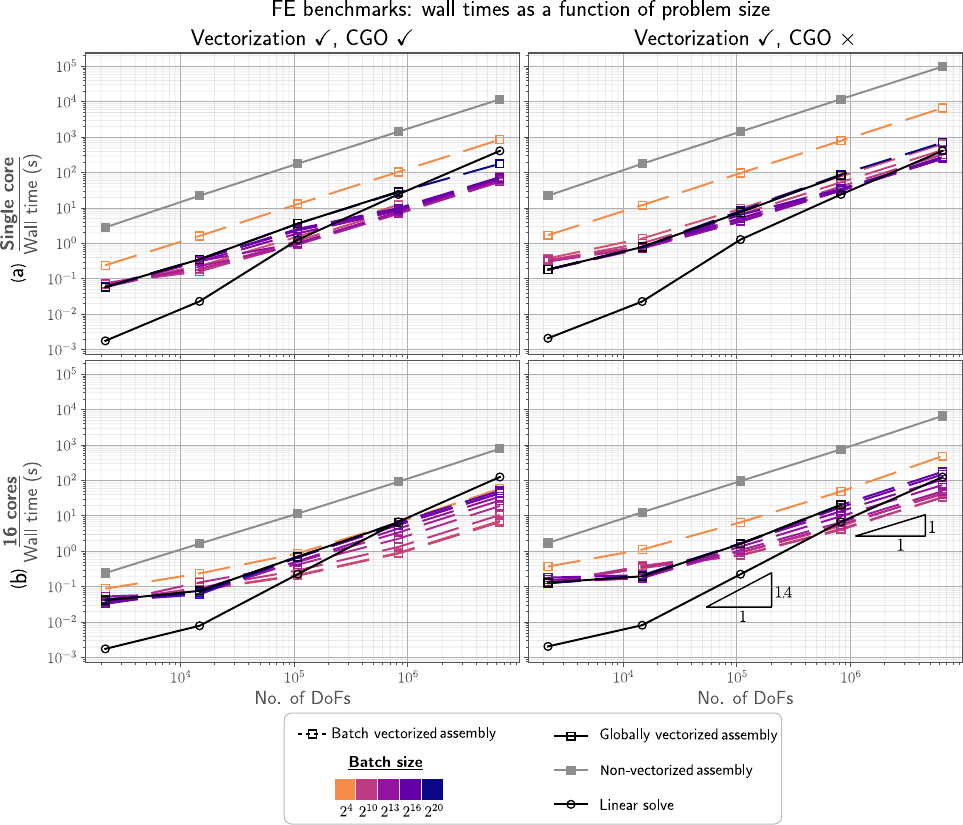}
	\end{center}
	\caption{\textbf{Scaling of assembly and solver wall times with problem size:} (left column) the combined effects of vectorization and CGO and (right column) vectorization without CGO are displayed in (a) the single core context and (b) multicore context. Assembly retains linear scaling with increasing degrees of freedom, while solver complexity grows superlinearly.}
	\label{fig:wall-times}
\end{figure}

\vspace{12pt}
The performance improvements due to vectorization, batching, and CGO are further contextualized in \fig~\ref{fig:wall-times} by comparing the wall time for solving the  linear system of equations with that of assembling the same linear system when using global-vectorization and batch-vectorization for various batch sizes as a function of the number of degrees of freedom (DoFs).

\textbf{Linear (optimal) scaling of assembly wall time with the number of DoFs is maintained when assembly is globally- or batch-vectorized} (\fig~\ref{fig:wall-times} all subfigures).
Furthermore, this scaling is maintained in the single core and multicore contexts, both for the CGO and non-CGO cases.
By contrast, solving the linear system of equations using the conjugate gradient method with a Jacobi preconditioner has an empirically derived computational complexity of approximately $\bigO\fOf{\nDofs^{1.4}}$, where $\nDofs$ is the number of DoFs.
Hence, with our vectorized FE assembly approach, there will be some problem size for which solving the linear system of equations becomes the dominant computational bottle-neck.
However, in these experiments, even for problem sizes of 6,440,067 DoFs, which is larger than is used in many practical applications, the wall time for non-vectorized (batch size of 1) non-CGO assembly is more than 100 times larger than that of the solving the linear system of equations (\fig~\ref{fig:wall-times} (a) right).

\textbf{Batched-vectorization and CGO alleviates assembly as the computational bottle-neck for problems with more than 100,000 DoFs} (\fig~\ref{fig:wall-times} (a) left).
Although vectorization does not alter the linear (ideal) scaling behavior of assembly, it reduces the assembly time by a machine- and batch size dependent coefficient (as discussed in \sect~\ref{sec:res-vec-batch} and \fig~\ref{fig:mp-scaling}), thus resulting in the transition of the computational bottle-neck from being assembly to being the solving of the linear system of equations at a much smaller problem size.
Hence, to speed up the FE simulation overall further, it would be more effective to put effort into speeding up the linear solver as it is now the dominant computational bottle-neck for the majority of problem sizes of practical relevance.
\clearpage

\subsection{Distributed memory parallelization benchmarks}
\label{sec:res-mpi}

\begin{figure}[h]
	\begin{center}
		\includegraphics[width=\linewidth]{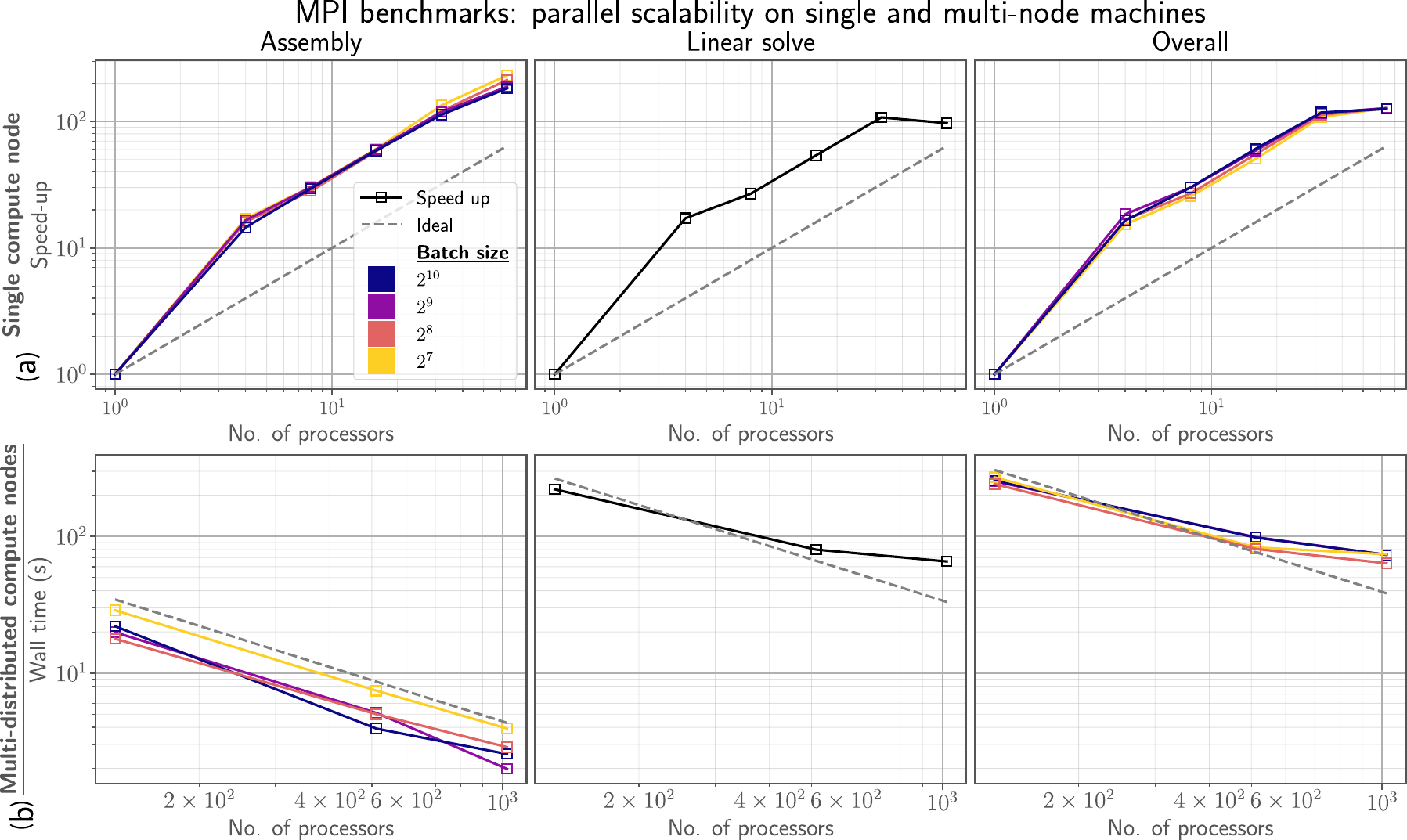}
	\end{center}
	\caption{\textbf{Strong scaling behavior of our developed FE solver under MPI parallelization.} For (left) assembly, (middle), linear solve, and (right) the overall simulation we report as a function of the number of processors used in the computation (a) the speed-up for a fixed problem size of 6,440,067 DoFs on a single compute node and (b) the wall time for a fixed problem size of 50,923,779 DoFs on a machine with distributed compute nodes. Moreover, values are reported for a range of batch sizes. }
	\label{fig:scaling}
\end{figure}

To evaluate parallelization performance of the batch-vectorized assembly algorithm (\alg~\ref{alg:bat-vec-asm}) and surrounding FE solver, we evaluate the \textit{strong scaling} in \fig~\ref{fig:scaling}, i.e. the speed-up of the wall time when using multiple processors relative to the single processor wall time, for both the single compute node and multi-compute nodes.\footnote{These computational experiments were run on the Delft Blue Supercomputer \cite{DHPC2024}. See \tab~\ref{tab:deets} in \ref{sec:mach-deets} for details of the CPUs used on these compute nodes.}
In a single compute node, data exchange between multiple processors is faster because all processors share the same memory.
In contrast, in a multi-node setup, data must be communicated over interconnects, which can introduce latency and slow down performance.

We focus only on the batch-vectorized case here since, for larger problems, the globally-vectorized algorithm introduces significant RAM requirements as discussed in \sect~\ref{sec:fe-asm-and-batch}.
Moreover, we note that our results in \sect~\ref{sec:res-fe-vec-batch} show that batch-vectorized assembly outperforms globally vectorized assembly with respect to compute time as well as RAM requirements.

\textbf{The batch-vectorized assembly algorithm and FE solver displays super-linear strong scaling in the single node case} (\fig~\ref{fig:scaling} (a)); i.e., the observed speed-up is greater than the ideal linear upper-bound proposed by Amdahl's law \cite{Amdahl1967_298e}.
Super-linear scaling typically results from the fact that each CPU has its own L1 and L2 cache \cite{Hager2011_244d}.
As more CPUs are made available to the program, a larger total amount of CPU cache is also made available ultimately reducing the amount of latency that is introduced due to main memory reads.

\textbf{The batch-vectorized assembly algorithm and FE solver displays good strong scaling behavior on as many as 1,024 cores on a multi-node setup} (\fig~\ref{fig:scaling} (b)).
In this case, we no longer observe super-linear scaling, which can be attributed to  communication latency between multiple compute nodes via interconnects.
We note, however, that this is not a feature of the program but rather the system on which the program is run.
At the same time,  \textbf{the scaling behavior of the batch-vectorized assembly algorithm outperforms that of the state-of-the-art linear solver implementation provided by PETSc} \cite{Balay2025_3565} (\fig~\ref{fig:scaling} (a) and (b), left and middle).
This ensures that assembly remains at least comparable to, if not more scalable than, the linear solve and is therefore unlikely to become the scaling bottleneck in large-scale FE simulations.

\section{COMMET: Demonstration of performance on large-scale practical FE simulations}
\label{sec:demo}

COMMET is not merely an academic exercise to showcase the effects of vectorization, batching, and CGO on simple structured-mesh benchmarks.
Instead, it delivers full finite element functionality for a wide range of solid mechanics problems, including support for unstructured meshes, three-dimensional field definitions, and diverse 3D boundary conditions.

To illustrate the capability of COMMET beyond canonical benchmarks, we perform a patient-specific simulation of human heart inflation under physiological loading conditions as an example problem.
The geometry was reconstructed from high-resolution magnetic resonance images of a healthy 44-year-old male subject (178 cm, 70 kg) \cite{Peirlinck2021_6f74}.
Diastolic filling was simulated by applying endocardial pressures of 8 mmHg and 4 mmHg in the left and right ventricles, respectively, representing physiologic end-diastolic states \cite{peirlinck2019kinematic}.
A pericardial constraint was imposed through Robin-type boundary conditions \cite{Arostica2025_5569}.
Myocardial tissue was modeled with a MICNN-based NCM, parameterized to reflect average biaxial stiffness of human myocardium \cite{Sommer2015_cc75}.

\begin{figure}[h]
	\center
	\includegraphics[width=.99\linewidth]{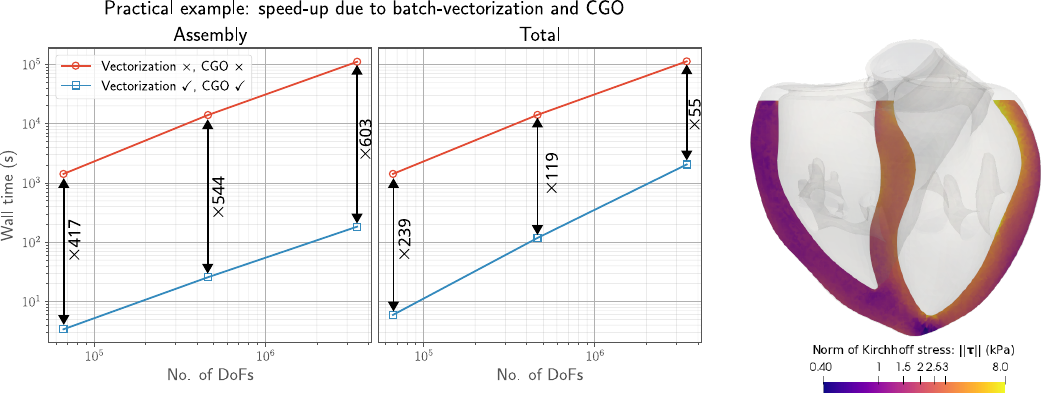}
	\caption{\textbf{COMMET functionality showcase.} Using NCMs to simulate the diastolic filling of a patient-specific heart.
		We compute the end-diastolic deformation and stretch in the human heart, for which the myocardium was modelled using a MICNN-based NCM.
		The wall times for vectorized CGO implementation are compared to that of non-vectorized non-CGO implementations for assembly (left) and the simulation overall (middle) at three different mesh refinement levels.
		The resulting Kirchhoff stress distribution is shown on the right.
	}
	\label{fig:real}
\end{figure}

\fig~\ref{fig:real} (right) depicts the resulting end-diastolic stress distributions.
To once again demonstrate the speed-up due batch-vectorization and CGO, the same problem was solved using a non-vectorized non-CGO implementation and batch-vectorized CGO (batch size of 512) implementation.
Moreover, the problem was solved for three levels of mesh refinement resulting in problem sizes of $66\,234$, $462\,474$, and $3\,436\,362$ degrees of freedom.
\fig~\ref{fig:real} (left) and (right) show that batch-vectorization and CGO result in reducing the time for assembly by a factor 417--603 times and reducing the time for the total simulation by a factor of 239--55.
Hence, it is clear that the computational gains of vectorization, batching, and CGO extend beyond synthetic benchmarks to complex, real-world geometries.
Such efficiency enables high-fidelity simulations in solid mechanics at scales that were previously impractical, and creates opportunities for large-scale studies requiring repeated solves, parameter sweeps, or uncertainty quantification.

\section{Conclusion}
\label{sec:con}

In this work, we have presented COMMET, a scalable and performant finite element solver designed to accelerate computationally intensive constitutive updates. Neural constitutive models (NCMs) represent an extreme but illustrative case, as their large computational graphs make repeated evaluations of stress and stiffness particularly costly, yet the same bottlenecks arise for many advanced material models in nonlinear solid mechanics. Our contributions are threefold: (i) globally and batch-vectorized assembly algorithms that restructure the traditional update loop to allow simultaneous evaluation of many material points, (ii) compute-graph-optimized derivatives that replace automatic differentiation and provide exact gradients at a fraction of the runtime and memory cost, and (iii) full compatibility with distributed-memory parallelism via MPI to ensure scalability across multiple compute nodes.

Extensive computational experiments demonstrated speed-ups exceeding three orders of magnitude in constitutive evaluations relative to traditional non-vectorized AD-based implementations, with roughly two orders of magnitude attributable to batch-vectorization and an additional 2–10× improvement from CGO depending on batch size. Batch-vectorization consistently outperformed global vectorization, exhibited an optimal batch size balancing cache efficiency with memory usage, and reduced RAM requirements compared to global vectorization. Parallel benchmarks showed superlinear scaling on single nodes and robust scaling to thousands of cores across distributed nodes, ensuring that assembly no longer constitutes the limiting factor in large-scale FE analyses.

Although our demonstrations focused on NCMs, the framework is not restricted to them: the same strategies apply wherever loop-based constitutive updates dominate runtime, from sophisticated anisotropic plasticity to multiscale homogenization.
COMMET therefore lays a strong foundation not only for the practical deployment of NCMs but also for accelerating high-fidelity FE simulations more broadly across solid mechanics.
\RV{Future work will target automatic batch-size tuning, support for history-dependent materials, multiphysics extensions, and GPU implementation}.
Through the open-source release of COMMET, we invite the community to adopt, extend, and accelerate both neural and conventional constitutive models in computational mechanics.

\appendix

\newpage

\clearpage

\section{Hyperelasticity formulations}\label{sec:hyperelasticity}

\subsection{Material and spatial stiffnesses}\label{sec:mat-spatial}
To allow for a generic interface in our code, we elect for defining the strain energy density $\en$ as a function of the deformation gradient $\F$ instead of e.g. the left or right Cauchy-Green tenors, $\C$ or $\B$, respectively.
However, taking first and second derivatives of $\en$ with respect to $\F$ yields the non-symmetric first Piola-Kirchhoff stress $\P$ and associated fourth order stiffness tensor $\CC^{\P}$, respectively,
\begin{align}
	P_{iJ} & := \pd{\en}{F_{iJ}}\,, & \CC^{\P}_{iJkL} & := \pdd{\en}{F_{iJ}}{F_{kL}}\,.
\end{align}
The lack of symmetry in these tensors preclude the usage of Voigt notation representations in code which would allow for significantly more performant tensor operations, particular in the case of the fourth order stiffness tensor.
Hence, to allow for the performance gains provided by Voigt notation, we transform these stress and stiffness tensors into the symmetric spatial counterparts.
We use the well-known push-forward operation to obtain the symmetric Kirchhoff stress,
\begin{equation}
	\tau_{ij} = \pd{\en}{F_{iJ}}F_{jJ}\,.
\end{equation}

Obtaining the transformation for the stiffness tensor is less trivial.
We start by noting
\begin{equation}
	\CC^{\P}_{iJkL} = \pd{}{F_{kL}}\brac{F_{iI}S_{IJ}} = \delta_{ik} S_{JL} + F_{iI}\CC_{IJKL}F_{kK}\,,
	\label{eq:p-stiffness}
\end{equation}
where $\S=\F^{-1}\P$ is the second Piola-Kirchhoff stress, $\CC=2\pd{\S}{\C}$ is the material stiffness tensor, and we have used the identity
\begin{equation}
	\pd{\bullet}{F_{kL}} = \pd{\bullet}{C_{IJ}}\brac{\delta_{LI}F_{kJ} + F_{kI} \delta_{JL}}\,.
\end{equation}
The spatial stiffness tensor is related to the material stiffness tensor by the well-known push forward operation
\begin{equation}
	\cc_{ijkl} = \CC_{IJKL}F_{iI}F_{jJ}F_{kK}F_{lL}\,.
	\label{eq:mat-spa-relation}
\end{equation}
By rearranging \eq~\eqref{eq:p-stiffness} and substituting into \eq~\eqref{eq:mat-spa-relation} we obtain
\begin{align}
	\cc_{ijkl} = F_{jJ}\brac{\CC^{\P}_{iJkL} - \delta_{ik} S_{JL} }F_{lL} = F_{jJ}\CC^{\P}_{iJkL}F_{lL} - \delta_{ik} \tau_{jl} \,,
\end{align}
where we have used $\btau = \F\S\F^T$.
Hence, the necessary transformations for obtaining the Kirchhoff stress and spatial stiffness when defining $\en$ in terms of $\F$ are
\begin{align}
	\tau_{ij} & = \pd{\en}{F_{iJ}}F_{jJ}\,, & \cc_{ijkl} & = F_{jJ}\pdd{\en}{F_{iJ}}{F_{kL}}F_{lL} - \delta_{ik}\tau_{jl}\,.
\end{align}

\subsection{Kinematic layers and derivatives for compute graph optimization}\label{sec:kinnematic-layers}
In most cases, the kinematic scalars used as inputs to the inner layer can be obtained from the right Cauchy-Green tensor.
Hence, using the chain-rule as discussed in \sect~\ref{sec:intro-cgo} and applying the relevant push forward operations yields the following expressions for the Kirchhoff stress and the spatial stiffness tensor:
\begin{align}
	\ta        & = 2\sum_m \pd{\actN}{\actK_m}\underbrace{\pd{\actK_m}{C_{IJ}}F_{iI}F_{jJ}}_{G^{m}_{ij}}\,\label{eq:stress-cgo}                                                                                                                                                                                          \\
	\cc_{ijkl} & = 4\sum_{m,n}\pdd{\actN}{\actK_m}{\actK_m}\underbrace{\pd{\actK_m}{C_{IJ}}F_{iI}F_{jJ}}_{G^{m}_{ij}}\underbrace{\pd{\actK_n}{C_{KL}}F_{kK}F_{lL}}_{G^{n}_{kl}}+4\sum_m \pd{\actN}{\actK_m}\underbrace{\pdd{\actK_m}{C_{IJ}}{C_{KL}}F_{iI}F_{jJ}F_{kK}F_{lL}}_{\GG^{m}_{ijkl}}\,.\label{eq:stiffnes-cgo}
\end{align}
Here, we have grouped the derivatives of the kinematic layer along with the deformation gradients resulting from the push-forward operations and define these as
\begin{align}
	\G^{m} & := \F \pd{\actK_m}{\C} \F ^{T}\,, & \GG^{m}_{ijkl} & := \pdd{\actK_m}{C_{IJ}}{C_{KL}}F_{iI} F_{jJ} F_{kK} F_{lL}\,.\label{eq:second-and-fourth}
\end{align}
Hence, we can determine the tensors $\G^{m}$ and $\GG^{m}$ for each kinematic scalar independently of the inner network used in the NCM.
Once, the first and second derivatives of the inner network are known, they can be combined with the corresponding $\G^{m}$ and $\GG^{m}$ to obtain the stress and stiffness tensors according to \eqs\eqref{eq:stress-cgo} and \eqref{eq:stiffnes-cgo}, respectively.
We elect for using the tensors defined in \eq~\eqref{eq:second-and-fourth} as opposed to, say $\pd{\actK_m}{\F}$ and $\pdd{\actK_m}{C_{iJ}}{F_{kL}}$, as they are symmetric by construction and conveniently allow for the use of Voigt notation.
We now proceed to present expressions for this second- and fourth-order for the case of standard and isochoric invariants, while noting that this approach can be applied similarly to the case of principal stretches.

\subsubsection{Invariants}
The standard invariants, and the corresponding second and fourth order tensors as defined in \eq\ref{eq:second-and-fourth} are given by
	{\footnotesize
		\begin{align}
			\iI     & = \tr{\C}\,,                                     & \G^{1}    & := \B\,,                                     & \GG^{1}    & := \OO\,, \label{eq:standard-start}                                                                                     \\
			\iII    & = \frac{1}{2}\brac{\tr{\C}^{2} - \tr{\C^{2}}}\,, & \G^{2}    & :=  \B\tr{\B} - \B^{2} \,,                   & \GG^2      & := \B\otimes\B - \B\stimes\B\,,                                                                                         \\
			\iIII   & = \det{\C}\,,                                    & \G^3      & := \det{\B} \I \,,                           & \GG^3      & := \det{\B}\brac{\I\otimes\I - \I\stimes\I}\,,                                                                          \\
			\iIVind & = \StrucVeci\cdot \C \StrucVecj\,,               & \G^{4,ij} & := \sym{\strucVeci\otimes \strucVecj}\,,     & \GG^{4,ij} & := \OO\,,                                                                                                               \\
			\iVind  & = \StrucVeci\cdot \C^{2}\StrucVecj\,,            & \G^{5,ij} & :=  2\sym{\strucVeci\otimes \B\strucVecj}\,, & \GG^{5,ij} & := \B\stimes\sym{ \strucVeci\otimes\strucVecj} +  \sym{\strucVeci\otimes\strucVecj}\stimes \B\,.\label{eq:standard-end}
		\end{align}
	}
Here, $\StrucVeci$ is the $i^{\text{th}}$ structural vector,
$\strucVeci:=\F\StrucVeci$ is the current configuration counterpart of $\StrucVeci$,
$\sym{\bullet}:=\frac{1}{2}\brac{\bullet + \bullet^{T}}$ is the symmetric part of a tensor, and the various tensor products are defined as follows:
\begin{align}
	\brac{\a \otimes \b}\bm{c} & = a_i b_jc_j                                                                                  \\
	\A \otimes \B              & = A_{ij} B_{kl} \e_i\otimes \e_j\otimes \e_k\otimes \e_l\,,                                   \\
	\A \stimes \B              & = \frac{1}{2} \brac{A_{ik} B_{jl}+ A_{il} B_{kj}} \e_i\otimes \e_j\otimes \e_k\otimes \e_l\,,
\end{align}
where $\e$ denotes a basis vector.

\subsubsection{Isochoric invariants}
Many hyperelastic materials exhibit behavior that is far stiffer in volumetric deformation than in isochoric (volume-preserving) deformation.
For this reason, it is common to multiplicatively decompose the deformation gradient into an isochoric $\FIso$ and volumetric part $\FVol$; that is,
\begin{equation}
	\F=\FIso\FVol\,,\qquad \FIso := J^{-1/3}\F\,,\qquad \FVol := J^{1/3}\I\,,\qquad J = \det{\F}\,.
\end{equation}
Here, $J$ is the (volumetric) Jacobian and $\iso{\bullet}$ and $\vol{\bullet}$ denote the isochoric and volumetric parts of $\bullet$, respectively.
The strain energy density function is then postulated in terms of the isochoric invariants,
\begin{align}
	\ImIso     & :=  \Im\IThree^{\eIsom} \qquad \text{for } m \in \invSetIso\,,                                                \\
	\invSetIso & := \set{1, 2, \fOf{4,ij},\, \fOf{5, ij},\, | \, i,j\in\brac{1,\, \nsv}}\,,                                    \\
	\eIsom     & := \begin{cases}
		                -2/3 & \text{if } m\in \set{2, \, \fOf{5, ij}\, | \, i,j\in\brac{1,\, \nsv}}\,, \\
		                -1/3 & \text{otherwise.}
	                \end{cases}
\end{align}
The corresponding second- and fourth-order tensors defined in \eq~\eqref{eq:second-and-fourth} are then given by
\begin{align}
	\G^{m}  & = \GIso^{m}+\eIsom\ImIso\I\,,                                                                                    \\
	\GG^{m} & = \GGIso^m + \eIsom\brac{\I\otimes \GIso^m + \GIso^m \otimes \I + \ImIso\brac{\eIsom\I\otimes\I-\I\stimes\I}}\,,
\end{align}
where $\GIso^m$ and $\GGIso^m$ are the isochoric versions of the corresponding terms in \eqs\eqref{eq:standard-end}--\eqref{eq:standard-end}, i.e.
	{\footnotesize
		\begin{align}
			\GIso^{1}    & := \BIso\,,                                        & \GG^{1}       & := \OO\,,                                                                                                          \\
			\GIso^{2}    & :=  \BIso\tr{\BIso} - \BIso^{2} \,,                & \GGIso^2      & := \BIso\otimes\BIso - \BIso\stimes\BIso\,,                                                                        \\
			\GIso^{4,ij} & := \sym{\strucVecIsoi\otimes \strucVecIsoj}\,,     & \GGIso^{4,ij} & := \OO\,,                                                                                                          \\
			\GIso^{5,ij} & :=  2\sym{\strucVecIsoi\otimes \B\strucVecIsoj}\,, & \GGIso^{5,ij} & := \BIso\stimes\sym{ \strucVecIsoi\otimes\strucVecIsoj} +  \sym{\strucVecIsoi\otimes\strucVecIsoj}\stimes \BIso\,.
		\end{align}
	}

At the same time, strain energy due to volumetric changes are modelled using $J$, for which the corresponding second- and fourth-order tensors are
\begin{align}
	\G & = \frac{J}{2} \I\,, & \GG & = \frac{J}{4}\brac{I\otimes\I - 2 \I \stimes \I}\,.
\end{align}

\newpage
\section{Inner neural constitutive networks} \label{sec:inner-networks}
Here we briefly present the architectures for several NCMs from literature including CANNs \cite{Linka2023_6533,Peirlinck2024_f38b,Peirlinck2024_34b9,Peirlinck2024_55e5}, MICNNs \cite{Klein2022_3243,Thakolkaran2022_37f6,Amos2017_773c,Jadoon2025_48c6}, and ICKANs \cite{Thakolkaran2025_6b61,Liu2025}.
The presentations here are kept brief and readers are referred to the original publications for detailed treatments.
Additionally, we provide analytical expressions for the first and second derivatives of CANNs and MICNNs as didactic examples for usage in CGO.
These expressions can be similarly derived for ICKANs and other NCM inner networks.

\subsection{Constitutive artificial neural networks (CANNs)}
CANNs \cite{Linka2023_6533,Peirlinck2024_f38b,Peirlinck2024_34b9,Peirlinck2024_55e5} have a tree-like architecture that is expressed mathematically as
\begin{equation}
	\actN\fOf{\bm{\actK}} = f_2 \circ f_1 \circ f_0 \fOf{\bm{\actK}} = \sum_{m\in\invSet}\sum_{k=1}^{n}w_{2,k,m}f_{2}\fOf{f_{1}\fOf{f_{0}\fOf{\actK_m; w_{0,k,m}}; w_{1,k,m}}}\,,
	\label{eq:cann-def}
\end{equation}
where, $\invSet$ is an enumeration of the kinematic scalars used as input to the network, $w_{i,k,m}$ $i=1,2,3$, $k=1,...,n$ $k\in\invSet$ are trainable weights, and
\begin{align}
	f_0 & = \begin{cases}
		        \fOf{\circ}                \\
		        \left<\circ\right>         \\
		        \left\vert\circ\right\vert \\
		        \vdots
	        \end{cases}                 &
	f_1 & = \begin{cases}
		        \fOf{\circ}^{1} \\
		        \fOf{\circ}^{2} \\
		        \fOf{\circ}^{3} \\
		        \vdots
	        \end{cases}                            &
	f_2 & = \begin{cases}
		        w_1\fOf{\circ}             \\
		        \exp{w_1\fOf{\circ}}-1     \\
		        -\ln\fOf{1-w_1\fOf{\circ}} \\
		        \vdots
	        \end{cases}\,.
	\label{eq:cann-fs}
\end{align}
Following \cite{Peirlinck2024_55e5}, we obtain the first and second derivatives of $\actN$ using the chain-rule; these are,
\begin{align}
	\pd{\actN}{\nnInput_m}                  & = \sum_{k=1}^{n} w_{2,k,m}\pd{f_{2}}{\circ}\pd{f_{1}}{\circ}\pd{f_{0}}{\actK_{m}}\,,                                                                                                                                                                 \\
	\pdd{\actN}{\nnInput_{m}}{\nnInput_{m}} & = \sum_{k=1}^{n} w_{2,k,m}\brac{\brac{\pdd{f_{2}}{\circ}{\circ}\brac{\pd{f_{1}}{\circ}}^{2}+\pd{f_{2}}{\circ}\pdd{f_{1}}{\circ}{\circ}}\brac{\pd{f_{0}}{\nnInput_m}}^{2} + \pd{f_{2}}{\circ}\pd{f_{1}}{\circ}\pdd{f_{0}}{\nnInput_m}{\nnInput_m}}\,.
	\label{eq:cann-chain-rule}
\end{align}
Note that, due to the form of \eq\eqref{eq:cann-def}, $\pdd{\actN}{\nnInput_{m}}{\nnInput_{n}} = 0\,,\quad \forall m\neq n$.
In order to evaluate the derivatives in \eq\eqref{eq:cann-chain-rule} the first and second derivatives of the expressions in \eq\eqref{eq:cann-fs} are required; these are,
\begin{align}
	\pd{f_0}{\circ}         & = \begin{cases}
		                            1                                 \\
		                            \frac{1}{2}\fOf{1 + \sign{\circ}} \\
		                            \sign{\circ}                      \\
		                            \vdots
	                            \end{cases}                           &
	\pd{f_1}{\circ}         & = \begin{cases}
		                            1                \\
		                            2\fOf{\circ}^{1} \\
		                            3\fOf{\circ}^{2} \\
		                            \vdots
	                            \end{cases}                                                           &
	\pd{f_2}{\circ}         & = \begin{cases}
		                            w_1                          \\
		                            w_1\exp{w_1\fOf{\circ}}      \\
		                            \frac{w_1}{1-w_1\fOf{\circ}} \\
		                            \vdots
	                            \end{cases}                                \\
	\pdd{f_0}{\circ}{\circ} & = \begin{cases}
		                            0 \\
		                            0 \\
		                            0 \\
		                            \vdots
	                            \end{cases}                                                               &
	\pdd{f_1}{\circ}{\circ} & = \begin{cases}
		                            0                \\
		                            2                \\
		                            6\fOf{\circ}^{1} \\
		                            \vdots
	                            \end{cases}                                                           &
	\pdd{f_2}{\circ}{\circ} & = \begin{cases}
		                            0                                            \\
		                            w^{2}_1\exp{w_1\fOf{\circ}}                  \\
		                            -\frac{w^{2}_1}{\brac{1-w_1\fOf{\circ}}^{2}} \\
		                            \vdots
	                            \end{cases}
\end{align}

\subsection{Monotonic input convex neural networks (MICNNs)}
In short, input convext neural networks (ICNNs) \cite{Amos2017_773c,Klein2022_3243} are described by the following equations:
\begin{subequations}
	\label{eq:icnn}
	\begin{align}
		 &  & \z\lay{0}  & = \bm{\actK}\,, \label{eq:icnn-input}                                            \\
		 &  & \my\lay{k} & = \A\lay{k}\z\lay{k-1} + \B\lay{k}\z\lay{0}+\bias\lay{k}\,,\label{eq:icnn-mid-1} \\
		 &  & \z^{(k)}   & = \actFof{\my\lay{k}}\,,\label{eq:icnn-mid-2}                                    \\
		 &  & \actN      & = \A\lay{n}\z\lay{n-1} + \B\lay{n}\z\lay{0}\,.\label{eq:icnn-output}
	\end{align}
\end{subequations}
Here, $\bm{\actK}$ is the input to the network, $\z^{(n)}$ is the output of the network, $\actF$ is an activation function that is applied elementwise, $\bias\lay{k}$ are learnable bias vectors, and $\A\lay{k}$ and $\B\lay{k}$ are learnable weight matrices.
Additionally, \eqs\eqref{eq:icnn-mid-1} and \eqref{eq:icnn-mid-2} are applied iteratively for $k=1,\,...,\,n-1$; that is, for each hidden layer in the network.
Convexity of \eq\eqref{eq:icnn} in $\bm{\actK}$ is gauranteed if all values in $\A\lay{k}$ are non-negative for $k>0$ and $\actF$ is convex and monotonically non-decreasing.
Furthermore, convexity and non-decreasing monotonicity of \eq\eqref{eq:icnn} in $\bm{\actK}$ is gauranteed if all values in $\A\lay{k}$ and $\B\lay{k}$ are non-negative and $\actF$ is convex and monotonically non-decreasing.

The relative first derivatives of \eqs\eqref{eq:icnn-input}--\eqref{eq:icnn-output}, determined via use of the chain-rule, are as follows:
\begin{align}
	\pd{\actN}{\actK_m}                             & = A\lay{n}_{j}\pd{z\lay{n-1}_j}{\actK_m} + B_m\lay{n} \label{eq:icnn-grad-1}        \\
	\text{(no sum on j) }\pd{z\lay{n-1}_j}{\actK_m} & = \pd{\actF}{y\lay{n-1}_j}\pd{y\lay{n-1}_j}{\actK_m}\label{eq:icnn-grad-dz}         \\
	\pd{y\lay{n-1}_j}{\actK_m}                      & = A\lay{n-1}_{jk}\pd{z\lay{n-2}_k}{\actK_m} + B_{jm}\lay{n-1}\label{eq:icnn-grad-3}
\end{align}
Note that the expression for $\pd{z\lay{n-2}_j}{\actK_m}$ will be identical to that in \eq~\eqref{eq:icnn-grad-dz}, however ``$n-1$'' will be replaced with ``$n-2$''.
Hence, \eqs~\eqref{eq:icnn-grad-dz} and \eqref{eq:icnn-grad-3} can be applied recursively from $k=n$ to $k=1$, at which point the necessary derivatives are given by
\begin{align}
	\text{(no sum on j) }\pd{z\lay{1}_j}{\actK_m} & = \pd{\actF}{y\lay{1}_j}\pd{y\lay{1}_j}{\actK_m}\,, \\
	\pd{y\lay{1}_j}{\actK_m}                      & = A\lay{1}_{jm} + B_{jm}\lay{1}\,.
\end{align}

The second derivatives of \eqs\eqref{eq:icnn-input}--\eqref{eq:icnn-output}, determined via use of the chain-rule on \eqs\eqref{eq:icnn-grad-1}--\eqref{eq:icnn-grad-3}, are as follows:
\begin{align}
	\pdd{\actN}{K_m}{K_n}                             & = A\lay{n}_{j}\pdd{z\lay{n-1}_j}{K_m}{K_n}                                                                                                                           \\
	\text{(no sum on j) }\pdd{z\lay{n-1}_j}{K_m}{K_n} & = \pd{\actF}{y\lay{n-1}_j}\pdd{y\lay{n-1}_j}{K_m}{K_n} + \pdd{\actF}{y\lay{n-1}_j}{y\lay{n-1}_j}\pd{y\lay{n-1}_j}{K_m}\pd{y\lay{n-1}_j}{K_n}\label{eq:iccn-2grad-dz} \\
	\pdd{y\lay{n-1}_j}{K_m}{K_n}                      & = A\lay{n-1}_{jk}\pdd{z\lay{n-2}_{k}}{K_m}{K_n} \label{eq:iccn-2grad-end}\,.
\end{align}
Note that the recursive logic applies to \eqs\eqref{eq:iccn-2grad-dz} and \eqref{eq:iccn-2grad-end} in a similar manner to that applied to \eqs~\eqref{eq:icnn-grad-dz} and \eqref{eq:icnn-grad-3}.
Hence, the (M)ICNN, along with its first and second derivatives can be evaluated in one pass, without the use of automatic differentiation, as detailed in \alg\ref{alg:opt-icnn}.
There, $\odot$ denotes the Hadamard product.
\begin{algorithm}
	\caption{Evaluating an M(ICNN) along with its first and second derivatives with one pass}
	\label{alg:opt-icnn}
	\begin{algorithmic}[1]
		\State $ \z  \gets   \actK$
		\State $ \pd{\z}{\actK}  \gets \I$
		\State $\pdd{\z}{\actK}{\actK} \gets  \mathbb{O}$
		\For{$k = 1,\dots,n-1$}
		\State $\my \gets \A\lay{k}\z + \B\lay{k}\actK+\bias\lay{k}$
		\State $\pd{\my}{\actK} \gets \A\lay{k}\pd{\z}{\actK} + \B$
		\State $\pdd{\my}{\actK}{\actK} \gets \A\lay{k}\pdd{\z}{\actK}{\actK} $
		\State $\pdd{\z}{\actK}{\actK} \gets \pd{\actF}{\my}\odot\pdd{\my}{\actK}{\actK}+\pdd{\actF}{\my}{\my}\odot\pd{\my}{\actK}\otimes\pd{\my}{\actK}$
		\State $\pd{\z}{\actK} \gets \pd{\actF}{\my}\odot\pd{\my}{\actK}$
		\State $\z \gets \actFof{\my}$
		\EndFor
		\State $\actN \gets \A\lay{n}\z + \B\lay{n}\actK$
		\State $\pd{\actN}{\actK} \gets \A\lay{n}\pd{\z}{\actK} + \B\lay{n}$
		\State $\pdd{\actN}{\actK}{\actK} \gets \A\lay{n}\pdd{\z}{\actK}{\actK} $
		\State \textbf{Output:} $\actN,\, \pd{\actN}{\actK},\, \pdd{\actN}{\actK}{\actK}$
	\end{algorithmic}
\end{algorithm}

\subsection{Input-convex Kolmogorov-Arnold Networks (ICKANs)}
We first briefly introduce Kolmogorov-Arnold networks (KANs) \cite{Liu2025} and we then discuss how this architecture is altered to ensure input-convexity in-line with \cite{Thakolkaran2025_6b61}.
We note again that, the presentation brief and only included for completeness -- readers are referred to \cite{Liu2025,Thakolkaran2025_6b61} for more detailed treatments.
The architecture for KAN of $R$ layers is defined as follows:
\begin{align}
	\z\lay{0}  & = \bm{\actK}\,,                                                                                                                                             \\
	\bfz^{(r)} & = \Bigg[\sum_{j=1}^{n_{r-1}} \spli_{r-1,1,j}\Bigl(z^{(r-1)}_j\Bigr) \ , \ \dots, \ \sum_{j=1}^{n_{r-1}} \spli_{r-1,n_r,j}\Bigl(z^{(r-1)}_j\Bigr)\Bigg]^T\,, \\[1mm]
	\actN      & = \sum_{j=1}^{n_{R-1}} \spli_{R-1,1,j}\Bigl(z^{(R-1)}_j\Bigr)\,.
\end{align}
Here, $\spli_{i,j,k}$ are weighted trainable univariate splines, i.e.
\begin{align}
	\spli(x) & =  w_{s} \psi(x)
\end{align}
where $w_s$ is a trainable weight and $\psi$ is $k^{\text{th}}$-order a B-spline consisting of $n_b$ basis functions $B_{i,k}$ with control points $c_i$, i.e.
\begin{align}
	\psi(x)=\sum_{i=1}^{n_b} c_i B_{i,k}(x), \qquad \text{with} \quad \sum_{i=1}^{n_b} B_{i,k}(x)=1 \quad \text{for}\quad x\in [x_\text{min},x_\text{max}].
\end{align}
To define the $k^{\text{th}}$-order B-spline basis functions, we consider a set of $m_b=(k+n_b+1)$ knots $\{t_i\} ^{m_b}_{i=1}$
and apply De Boor's recursive algorithm \cite{de_Boor1972_6f9f} as follows:\\

\begin{adjustwidth}{12pt}{}
	\quad Zero-order basis function ($ k=0 $):
	\begin{equation}
		B_{i,0}(x) =
		\begin{cases}
			1, \quad \text{if } t_i \leq x < t_{i+1}, \\
			0, \quad \text{otherwise}.
		\end{cases}
	\end{equation}

	\quad Recursive definition for higher orders ($ k > 0 $):
	\begin{equation}
		B_{i,k}(x) = \frac{x - t_i}{t_{i+k} - t_i} B_{i,k-1}(x) + \frac{t_{i+k+1} - x}{t_{i+k+1} - t_{i+1}} B_{i+1,k-1}(x).
	\end{equation}
\end{adjustwidth}

For the special case of a uniform B-spline, the knots are equally spaced, i.e.,
\begin{equation}
	t_{i+2} - t_{i+1} = t_{i+1} - t_{i}, \quad \forall\ i \in [1, m_b-2].
\end{equation}
For a KAN to be input-convex, i.e. for a KAN to be an ICKAN, we require that the weights $w_s$ are positive and that the splines are convex and monotonically non-decreasing \cite{Thakolkaran2025_6b61}.
This is satisfied so long as the control points satisfy the following condition:
\begin{equation}
	c_{i+2} - c_{i+1} \geq c_{i+1} - c_{i} \geq 0, \quad \forall\  i \in [1, n_b-2].
\end{equation}

\newpage

\section{Data generation and NCM training}
\label{sec:data-gen-and-training}

When training all NCMs used in this work, we follow the EUCLID paradigm for unsupervised discovery of material behavior.
More specifically, we use the NN-EUCLID framework of Thakolkaran et al. \cite{Thakolkaran2022_37f6,Thakolkaran2025_6b61}.
In brief, this allows for the training of NCMs using full-field displacements and global reaction forces, both of which are physically obtainable from real experiments by using a combination of digital image correlation (DIC) and a load cell, i.e. stress measurements are not required.
Given the known displacements and reaction forces $R^{\beta, t}$ on $\beta=1,...,n_{\beta}$ constrained boundaries at $t=1,...,n_t$ time steps, the parameters $\params$ for a given NCM are obtained using
\begin{equation}
	\params  = \argmin_{\params} \sum_{t=1}^{n_t} \brac{\sum_{(I,i)\in D ^{\text{free}}}\fOf{r_{i}^{I,t}} ^{2} + \sum_{\beta=1} ^{n_\beta} \fOf{R ^{\beta, t} - \sum_{(I,i) \in D_{\beta} ^{\text{fix}}} r_{i}^{I,t}} ^{2}}
	\label{eq:loss}
\end{equation}
where $D^{\text{free}}$ is the set of tuples of nodes that are unconstrained in given direction and $D^{\text{fix}}_{\beta}$ is the set of nodes that are constrained in a given direction on boundary $\beta$.
Readers are referred to \cite{Thakolkaran2022_37f6,Thakolkaran2025_6b61} for the derivation of \eq\eqref{eq:loss}, however, in essence the sum over $D^{\text{free}}$ enforces that the discovered values for $\params$ result in the balance of linear momentum being satisfied for the given data and the sum over the boundaries $\beta=1,...,n_\beta$ enforces that the discovered values for $\params$ results in the observed reaction forces.

\begin{figure}
	\begin{center}
		\includegraphics[width=0.4\textwidth]{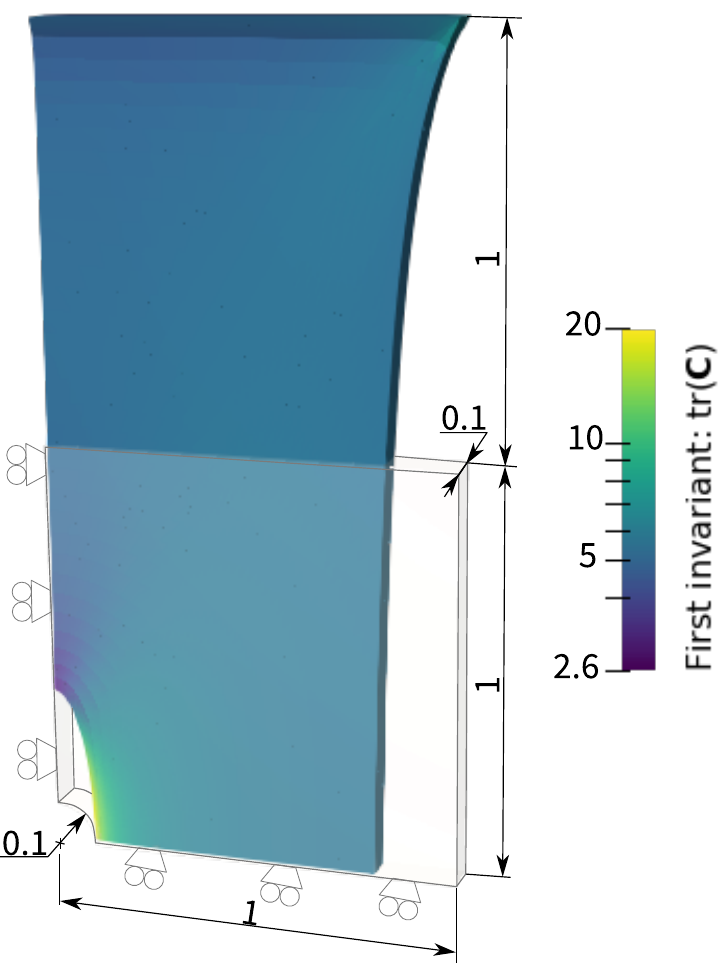}
	\end{center}
	\caption{\textbf{Specimen used for NCM training data generation.} Both reference and deformed configurations are shown.}
	\label{fig:test-specimen}
\end{figure}

For the purposes of this work, and without loss of generality, we generate synthetic data using a FE simulation.
We choose a Gent-Thomas material model \cite{Gent1958_2fcf}, defined by
\begin{equation}
	\en\fOf{\F} = 0.5(\iIi-3) + \log\fOf{\iIIi/3} + (J-1) ^{2}\,,
	\label{eq:gt}
\end{equation}
and specimen geometry and boundary conditions illustrated in \fig\ref{fig:test-specimen}.
The specimen consists of a $1\times 1$ square with of hole of radius $0.1$ in the bottom left corner that has been extruded by $0.1$.
Slider boundary conditions are applied on the left, bottom, and back of the specimen, while a unit of upwards displacement is applied to the top of the specimen.

{
    All NCMs made use of a kinematic layer that maps the deformation gradient to polyconvex terms as follows:
    \begin{equation}
        \actK\fOf{\F} = \begin{bmatrix}
            \iIi -3 & \iIIi^{3/2} - 3 ^{3/2} & (J-1) ^{2}
        \end{bmatrix} \,.
    \end{equation}

    The hyperparameters for the inner networks are provided in \tab~\ref{tab:hyperparams} and are comparable to those found in literature, e.g. \cite{Thakolkaran2022_37f6,Thakolkaran2025_6b61,Peirlinck2024_f38b}.
\begin{table}
	\caption{Hyperparameters used in NCMs}\label{tab:hyperparams}
	\begin{center}
        {
		\begin{tabular}[c]{l  l}
			Parameter &  Value \\
			\hline 
            \textit{MICNN \cite{Klein2022_3243,Thakolkaran2022_37f6,Amos2017_773c,Jadoon2025_48c6}: } &  \\
			\qquad Number of hidden layers & 3 \\
			\qquad Number of neurons in each hidden layer & 16 \\
			\qquad Total number of learnable parameters & 675 \\
			\hline
            \textit{CANN \cite{Linka2023_6533,Peirlinck2024_f38b,Peirlinck2024_55e5,Peirlinck2024_34b9}: } &  \\
            \qquad Exponents used in power layer & $\{1,2\}$ \\
            \qquad Terms used in function layer & $\{f(x)=x, \quad f(x)=\exp{x}\}$ \\
			\qquad Total number of learnable parameters & 33 \\
			\hline
            \textit{ICKAN \cite{Thakolkaran2025_6b61,Liu2025}:} &  \\
			\qquad Number of hidden layers & 2 \\
			\qquad Order of splines & 4 \\
			\qquad Number of grid points & 30 \\
			\qquad Total number of learnable parameters & 508 \\
			\hline
		\end{tabular}

        }
	\end{center}
\end{table}
}

\begin{figure}
	\begin{center}
		\includegraphics[width=0.6\textwidth]{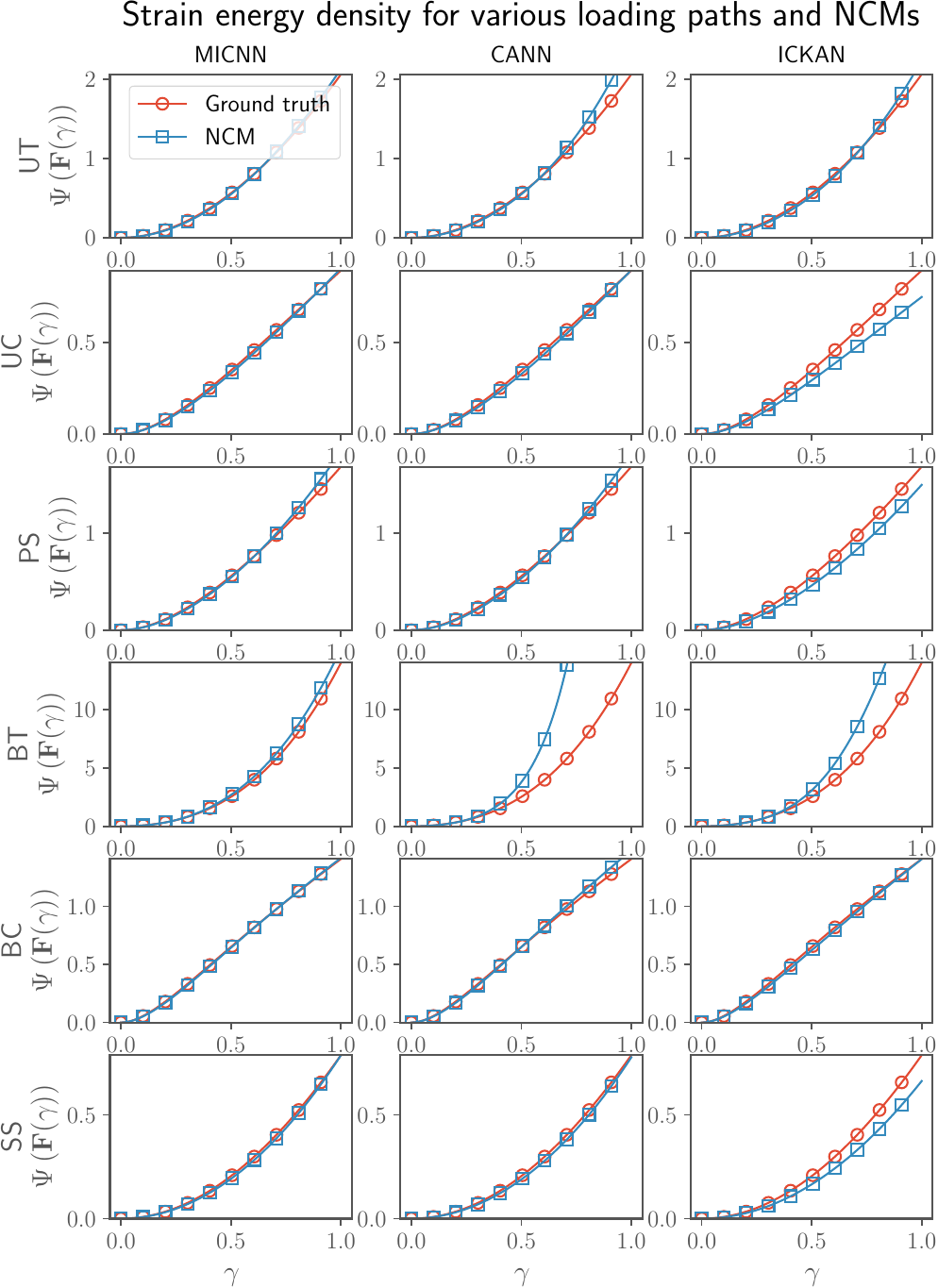}
	\end{center}
	\caption{\textbf{Comparison of trained NCM behavior with ground truth.} The NCMs are able to discover the ground truth behavior accurately in all cases apart from biaxial tension.}
	\label{fig:behave}
\end{figure}

The accuracy of the trained NCMs is evaluated by comparing the predicted strain energy density against that of the ground truth for six different loading paths, namely uniaxial tension (UT), uniaxial compression (UC), biaxial tension (BT), biaxial compression (BC), simple shear (SS), and pure shear (PS), defined, respectively, as follows:
\begin{equation}
	\begin{aligned}
		\F^{\text{UT}}\fOf{\gamma} & = \begin{bmatrix}
			                               1+ \gamma & 0 & 0 \\
			                               0         & 1 & 0 \\
			                               0         & 0 & 1
		                               \end{bmatrix}\,,                               &
		\F^{\text{UC}}\fOf{\gamma} & = \begin{bmatrix}
			                               \frac{1}{1+ \gamma } & 0 & 0 \\
			                               0                    & 1 & 0 \\
			                               0                    & 0 & 1
		                               \end{bmatrix}\,,                    &
		\F^{\text{BT}}\fOf{\gamma} & = \begin{bmatrix}
			                               1+ \gamma & 0         & 0 \\
			                               0         & 1+ \gamma & 0 \\
			                               0         & 0         & 1
		                               \end{bmatrix}\,,
		\\
		\F^{\text{BC}}\fOf{\gamma} & = \begin{bmatrix}
			                               \frac{1}{1+ \gamma } & 0                    & 0 \\
			                               0                    & \frac{1}{1+ \gamma } & 0 \\
			                               0                    & 0                    & 1
		                               \end{bmatrix}\,, &
		\F^{\text{SS}}\fOf{\gamma} & = \begin{bmatrix}
			                               1 & \gamma & 0 \\
			                               0 & 1      & 0 \\
			                               0 & 0      & 1
		                               \end{bmatrix}\,,                                  &
		\F^{\text{PS}}\fOf{\gamma} & = \begin{bmatrix}
			                               1+\gamma & 0                   & 0 \\
			                               0        & \frac{1}{1+ \gamma} & 0 \\
			                               0        & 0                   & 1
		                               \end{bmatrix}\,.
	\end{aligned}
\end{equation}
We note that these loading paths do not produce e.g. uniaxial tension in the typical sense as the $\tau_{22}$ and $\tau_{33}$ components of the resulting stress tensor will not in general be zero.
However, this is immaterial as the purpose is simply to compare the behavior of the trained NCMs to the ground truth for a small number of interpretable loading paths.
The resulting behavior for these loading paths is presented in \fig\ref{fig:behave}.
The NCMs are able to discover the ground truth behavior accurately in all cases apart from biaxial tension, the loading for which is outside of the training data.

\clearpage

\section{Machine details} \label{sec:mach-deets}
For completeness, we provide the details of the CPUs used in the computational experiments conducted in this work.
The details of the CPU in the workstation and on the HPC nodes are provided in \tab~\ref{tab:deets}.

\begin{table}[h]
	\caption{Details of CPUs used in computational experiments}\label{tab:deets}
	\begin{center}
		\begin{tabular}[c]{l| l | l}
			\hline
			                   & Workstation            & HPC node                    \\
			\hline
			Model name         & AMD Ryzen 9 7950X      & Intel(R) Xeon(R) Gold 6248R \\
			Core(s) per socket & 16                     & 24                          \\
			Socket(s)          & 1                      & 2                           \\
			CPU max MHz        & 5881                   & 4000                        \\
			CPU min MHz        & 400                    & 1200                        \\
			L1d cache          & 512 KiB (16 instances) & 32 KiB (48 instances)       \\
			L1i cache          & 512 KiB (16 instances) & 32 KiB (48 instances)       \\
			L2 cache           & 16 MiB (16 instances)  & 1 MiB (48 instances)        \\
			L3 cache           & 64 MiB (2 instances)   & 35.75 MiB (2 instances)     \\
			\hline
		\end{tabular}
	\end{center}
\end{table}

\clearpage

{
\section{The effect of model size} \label{sec:model-size}
The hyperparameter choices and model sizes considered in this work are comparable to those commonly reported in the literature \cite{Thakolkaran2022_37f6,Thakolkaran2025_6b61,Peirlinck2024_f38b}.
However, applications may require models with significantly more or fewer parameters than are typically employed.
Consequently, it is also of interest to assess how the performance gains arising from vectorization, batching, and CGO depend on the size of the NCM.

To this end, we repeat the material point benchmarks presented in \sect~\ref{sec:res-pt-cgo} for MICNNs with varying numbers of hidden layers and layer widths, thereby spanning a wide range of model sizes.
In particular, MICNNs are constructed with 2, 3, and 4, hidden layers and layer widths of 2, 4, 8, 16, 32, and 64.
In addition, to evaluate whether similar performance gains can be achieved for traditional constitutive laws, we repeat the material point benchmarks for a Gent–Thomas material model.

The resulting speed-ups are shown in \fig~\ref{fig:nn-size-speed-up} as a function of batch size for different NCM sizes.
Overall, speed-ups due to vectorization, batching, and CGO are larger for smaller NCMs.
We speculate that this is because the parameters for the NCM are held in cache during these computations. 
Hence, for bigger NCMs there is less cache memory available for the data on which the NCM operates.
However, since the contents of the CPU cache cannot be directly inspected, this hypothesis cannot be conclusively verified.
Unsurprisingly, the speed-up for the Gent-Thomas model is similar to that observed for NCMs with a small number of parameters, i.e. over three orders of magnitude.
Hence, batching and vectorization also leads to significant performance gains for traditional constitutive models.

\begin{figure}
\begin{center}
    \includegraphics[width=\textwidth]{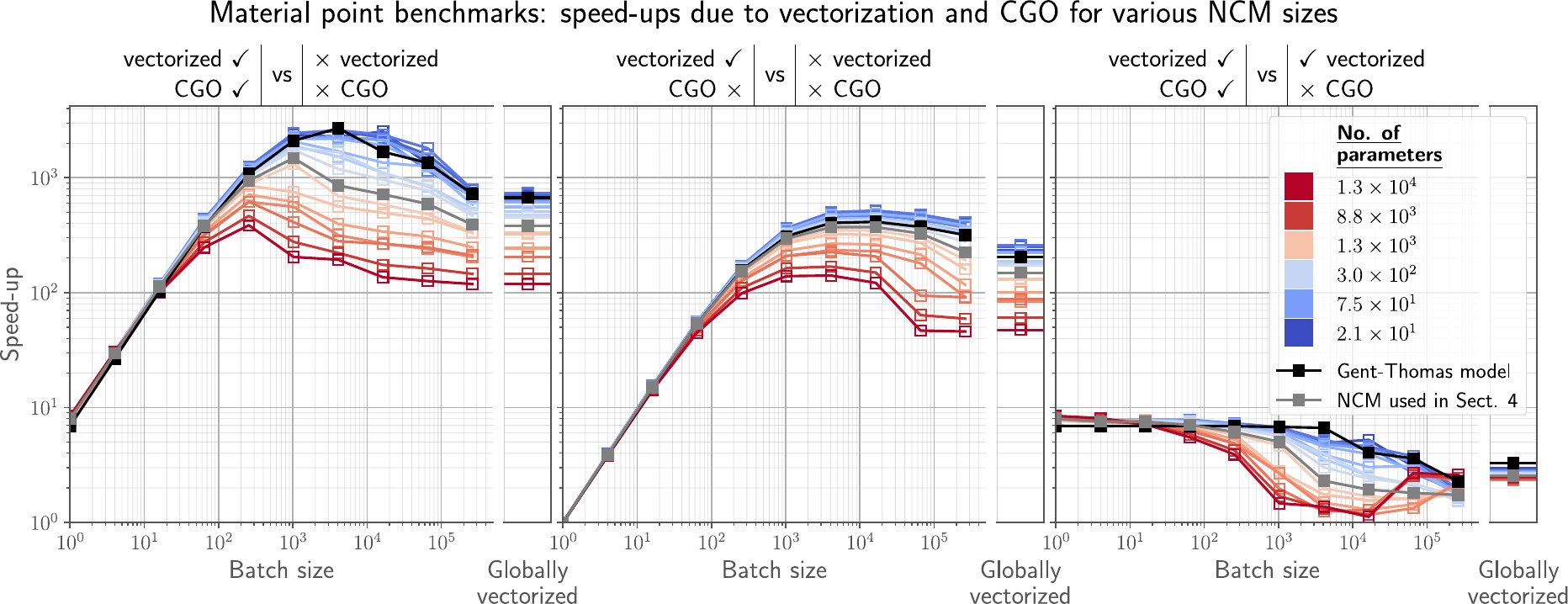}
\end{center}
\caption{\textbf{Effect of NCM size on speed-up due to batching, vectorization, and CGO:} (left) combined speed-ups from CGO and vectorization relative to a non-vectorized, non-CGO baseline; (middle) speed-ups due to vectorization only; and (right) additional speed-ups due to CGO only at different batch sizes. While increasing model size reduces the attainable speed-up, improvements of more than two orders of magnitude are still achieved for models substantially larger than those typically reported in the literature.}
\label{fig:nn-size-speed-up}
\end{figure}

Despite the observed decrease in speed-up with increasing model size, the performance gains remain substantial.
In particular, for NCMs with approximately 13{,}000 parameters, speed-ups exceeding two orders of magnitude are still obtained.
This is notable given that typical NCMs reported in the literature contain on the order of 1{,}000 parameters \cite{Thakolkaran2022_37f6}.

In addition to computational performance, memory requirements are an important practical consideration.
The RAM usage as a function of batch size and NCM size is therefore shown in \fig~\ref{fig:nn-size-ram}.
For the largest NCM considered, containing approximately 13{,}000 parameters, the total RAM usage remains below 1~GB for batch sizes smaller than $10^{4}$.
Notably, this range also encompasses the batch sizes that yield near-optimal speed-ups (\fig~\ref{fig:nn-size-speed-up}).
Such memory requirements are well within the capabilities of modern computing hardware.
Furthermore, a comparison of the left and right columns of \fig~\ref{fig:nn-size-ram} clearly indicates that the use of CGO reduces overall RAM consumption.

\begin{figure}
    \begin{center}
        \includegraphics[width=0.75\textwidth]{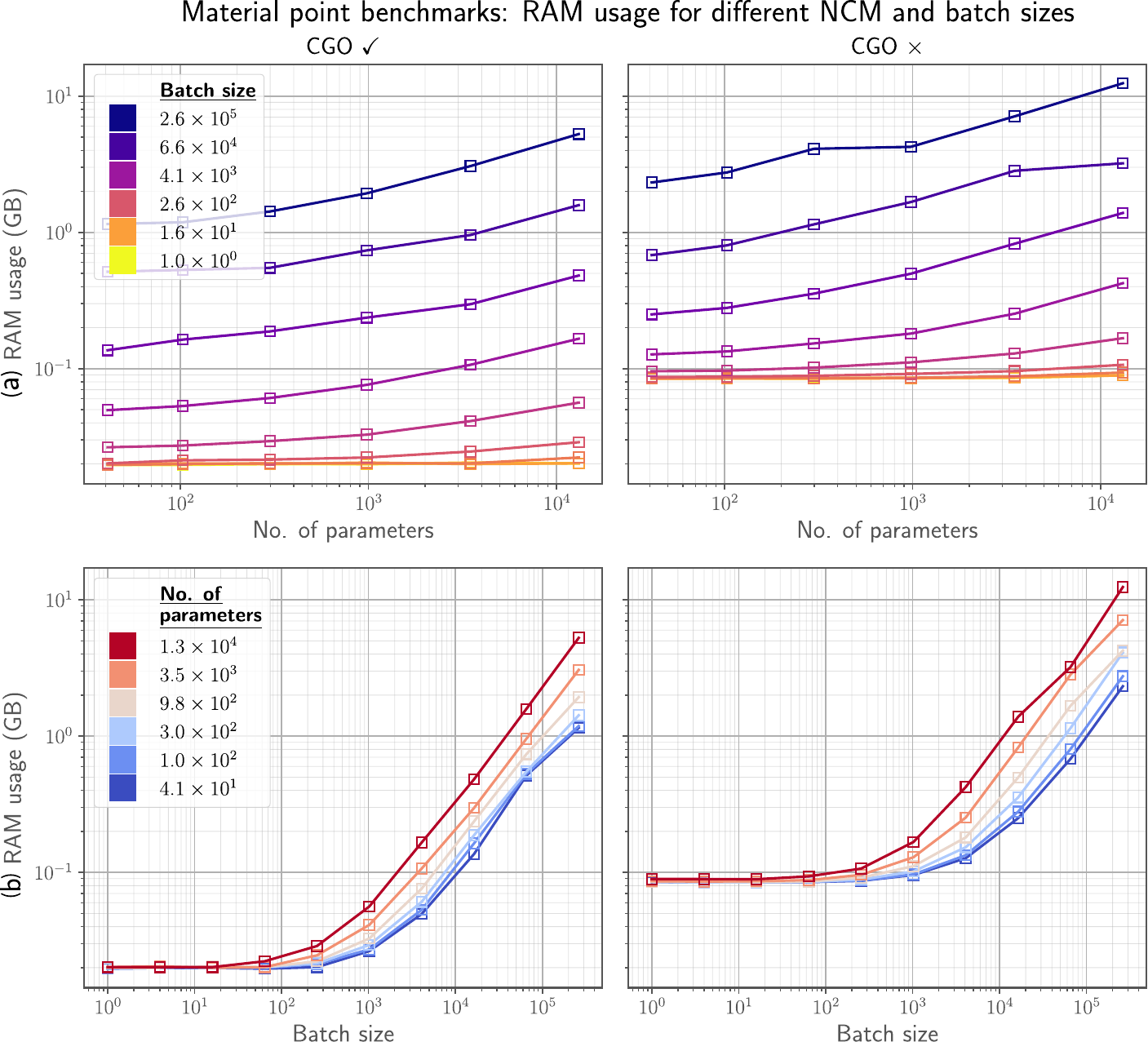}
    \end{center}
    \caption{\textbf{Effect of NCM size on RAM usage:} RAM consumption (left) with CGO and (right) without CGO, shown (a) as a function of the number of parameters and (b) as a function of batch size.}
    \label{fig:nn-size-ram}
\end{figure}

}

\newpage

\section*{Data and code availability}
\RV{The COMMET codebase is publicly available through \url{https://commet-code.github.io/} and \url{https://doi.org/10.5281/zenodo.17310682}.}

\bibliographystyle{elsarticle-num}
\bibliography{bib-file}

\end{document}